\documentclass[preprint,authoryear,12pt]{elsarticle}

\usepackage{amsmath,amssymb}
\usepackage{natbib}
\usepackage{amssymb,amsfonts,amsmath}
\usepackage{color}
\usepackage{array}
\usepackage{tabularx}
\usepackage{float}
\usepackage{graphicx}

\usepackage{enumitem}

\journal{Journal of Economic Dynamics and Control}

\begin{document}

\begin{frontmatter}

\title{Modeling urban housing market dynamics:
can the socio-spatial segregation preserve some social diversity?}

\author{}
\address{}
\author[label1]{Laetitia Gauvin\fnref{labnote}}
\author[label2]{Annick Vignes}
\author[label1,label3]{Jean-Pierre Nadal}
\fntext[labnote]{Corresponding author, laetitia.gauvin@lps.ens.fr}

\address[label1]{Laboratoire de Physique Statistique (LPS, UMR 8550 CNRS-ENS-UPMC-Univ. Paris Diderot), Ecole Normale Sup\'erieure, Paris, France}
\address[label2]{Equipe de Recherche sur les March\'es, l'Emploi et la Simulation (ERMES, EA 4441 CNRS-Paris II), University Paris II Panth\'eon-Assas, Paris, France}
\address[label3]{Centre d'Analyse et de Math\'ematique Sociales (CAMS, UMR 8557 CNRS-EHESS), Ecole des Hautes Etudes en Sciences Sociales, Paris, France}



\begin{abstract} 
Addressing issues of social diversity, we
introduce a model of housing transactions
between agents who are heterogeneous in their willingness to pay. A key
assumption is that agents' preferences for a
location depend on both an intrinsic attractiveness
 and on the social characteristics of the neighborhood.
The stationary space distribution of income is
analytically and numerically characterized.
The main results are that socio-spatial segregation occurs if -- and only if -- the social influence is strong enough, but even so, some social
diversity is preserved at most locations. Comparison with data on the Paris housing market shows that the results reproduce general
trends of price distribution and spatial income segregation.
\end{abstract}

\begin{keyword}
housing market model \sep income segregation \sep social
diversity \sep agent-based model
\\ {\bf J.E.L. codes:} R31; C61; C62; C63
\end{keyword}

\end{frontmatter}


\begin{flushright}
{\it Don't buy the house, buy the neighborhood} (Russian proverb)
\end{flushright}

 \section{Introduction}

\label{justif} People's choice of residential location and the way they are
distributed across cities matter, from both social and economic
points of view. This article seeks to explain, from the dynamics of price
formation in an urban housing market, how
individuals who are heterogeneous in their willingness to pay are distributed over a
city. It shows that, under certain conditions on social interactions, housing price
 formation can entail income segregation, even if
 a space of social diversity remains.

A large literature is concerned with evaluating the extent and impact
of housing price discrimination in education, housing and the labor
market. While \cite{Brueckner1999} explain that the relative location
of different income social groups depends on the spatial pattern of
amenities in a city, \cite{Gobillon2007}
 highlight the adverse labor-market outcomes due to spatial mismatch, with low-skilled inhabitants of the
inner-city suffering a greater distance to jobs and consequently a
higher level of unemployment. Understanding the formation of prices
in the real estate market and the way that prices are distributed over
space is clearly an important issue. Following the path opened by
\cite{Rosen1974}, most studies have focused on explaining prices
through hedonic estimations, showing how the price per square meter
can be influenced both by variables intrinsic to the apartment or
house and by extrinsic variables concerning the surrounding area and
its amenities. The role of these extrinsic variables has been
particularly explored. \cite{Baltagi2010} underline how much the
location influences the price of a dwelling. \cite{Ioannides2003},
\cite{Figlio2004}, \cite{Gravel2007} or \cite{Seo2009} emphasize the
importance of the quality and density of the neighborhood, the
reputation of nearby schools and the level of security. Following
\cite{Tiebout1956}, these authors point out that the decision of
where to live is based on families' preferences for the quality of
public services and amenities, particularly education. Thus, prices
on the real estate market vary with the quality of a bundle of public
services which are capitalized into housing prices.

One question underlined by the studies cited above is that of people's
preferences. Do they prefer to live with people who are richer than
they are, or poorer? In other words, what determines the social component of the
attractiveness of a location? The literature on well-being tends to
argue that people feel better when those around them are poorer (for
a detailed survey see \cite{Luttmer2005}). \cite{Clark2002} show
empirically that unemployed people are less unhappy
 when they live with other unemployed people. \cite{Goyal2010} explore
 the role of shifting social interactions: they use examples to illustrate
how poorer individuals lose while richer ones gain as we move from an
economically segregated society towards an integrated
society. In line with the literature on hedonic prices which
highlights the importance of the environment, we explore here the
consequences of the alternative hypothesis that individuals prefer to
live with people who are at least as rich as themselves. Indeed,
some preliminary studies suggest that the interplay between social
preferences and the preference for a high quality of local amenities
has more striking effects when the social preference is to live with
richer neighbors.

When the prices depend not only on the intrinsic characteristics of
the goods but also on the level of surrounding amenities, as
well as on some social preferences, the space is differentiated and a
social mismatch may result. 
 Different measures of segregation have been proposed. \cite{Alesina2011} measure
segregation of different ethnic, religious and linguistic groups
within the same country from quite an exhaustive data set covering several different countries. \cite{Cutler1999} show the influence of
legal barriers on social segregation, while \cite{Jenks1990} or
\cite{Cutler2008} evaluate the effect of segregation on the
socioeconomic performance of minorities. \cite{Echenique2007} develop
a spectral index of segregation related to the characteristics of
social interactions. This index is defined at the individual level
and is higher when the considered individual interacts with
segregated individuals. \cite{Ballester2010} propose a
random-walk-based segregation measure.
 For the present work, interesting results are obtained by considering simple
measures of multigroup segregation, which compare,
for a population in a given local neighborhood,
the observed distribution of a given feature (here income)
with the uniform distribution. A basic segregation index
in the line of \cite{ReardonFirebaugh2002} and \cite{Alesina2011} is proposed and compared to
the information-theoretic measure of \cite{Theil67}, which was introduced into
economics for measuring income heterogeneity.

The pioneering modeling work of
\cite{Schelling1971} describes residential
 segregation as emerging from social preferences alone.
 Since then, extensions have been proposed by several authors in order to
integrate
 a housing market.
In \cite{Benard2007}, the price of
 a house depends on an intrinsic component of the location,
 randomly allocated, and on the composition of the
 neighborhood. In  \cite{Fosset2003},
 the price only depends on an intrinsic component of the location uniformly distributed
over the city. However, the level of quality of locations does not
directly impact the choice made by individuals. In
\cite{Zhang2004}, the price varies according to the density of
occupation of the neighborhood, but does not take into account any
measure of house quality.
 In \cite{Benard2007}, the choice of a location
depends on the mean status of the neighborhood, the status being a
wealth-related quantity randomly allocated to the individuals. The
present work goes further, dealing with the individual attractiveness
of a location and its influence on prices: in our model, the
individuals choose a location according to its attractiveness, which
is a dynamic quantity depending on both the intrinsic characteristic
of the location and the (time-dependent) social composition (measured
by the levels of income) of the neighborhood. The prices then depend
on the
 attractiveness through the market dynamics.
 The framework introduced here can easily be adapted to make the
attractiveness
 reflect different characteristics of the locations and different types of social preferences.
It  also has the advantage of allowing for
 detailed mathematical analysis.

The present article focuses on spatial income segregation, leaving
aside all other features (such as ethnic characteristics) that may
also cause segregation. The model proposed here takes some
inspiration from \cite{Short} and
  \cite{Berestycki}, who model the evolution of the spatial distribution of crime in a city,
attributing to each location an attractiveness for illegal activity.
The originality here is that each agent attributes to each location a
specific level of
 attractiveness. This attractiveness results from a combination
  of an intrinsic or objective part, and a subjective part.
  The endogenous (subjective) attractiveness results from the
individuals' intrinsic social preferences (for neighbours with similar or higher incomes). The main
 assumptions are: (\emph{i}) people make decisions according to both
 their willingness to pay (WTP) and
their individual evaluation of the level of attractiveness of the
different locations; (\emph{ii}) buyers, who are heterogeneous in their WTP,
base their search for housing on the level of attractiveness of the
location of the dwelling; (\emph{iii}) agents are both buyers and
sellers; (\emph{iv}) the intrinsic attractiveness
 is spatially heterogeneous.

The results are demonstrated through mathematical analysis and then
empirically confirmed in the simulations and in the empirical
analysis using data on the Paris housing market. The analysis of
the stationary regime reached by the market dynamics is first performed
for an arbitrary, spatially heterogeneous, intrinsic attractiveness.
For illustrative purposes and to compare with the empirical data, the
analysis is then conducted in more detail for specific cases. First,
we consider the case of a monocentric city defined by an intrinsic
attractiveness which decreases with the distance from the
geographical center. This is a simple case motivated by the classical
Von Th\"{u}nen model and Alonso's study of land use (see
\cite{Alonso1964}), explaining how, generally, prices are higher in
the center of a city or a county. The fact that prices diffuse from
the center to the periphery (with higher prices in the center) has
more recently been highlighted in a regional context by
\cite{Clapp1995}, \cite{Meen1996}, \cite{Berg2002},
\cite{OIKARINEN2005} or \cite{Holly2010}. The monocentric case also
provides a first order approximation to the description of the
Paris housing market. Second, we consider a more complex structure of the
intrinsic attractiveness, allowing for a better comparison with
empirical data on transaction prices in Paris.

This article exhibits three important results.
The first concerns the mathematical analysis of the model. This analysis is first presented for an arbitrary intrinsic attractiveness. This provides general results. The results are then specified for the particular case of a monocentric city and this monocentric case is simulated. For the empirical validation, we consider a more complex structure of the intrinsic attractiveness which matches the Paris data quite well. Within this theoretical framework, we provide an analytic description of the spatial distributions of transaction prices and incomes in the city. We prove that, when the market is not saturated, social segregation occurs if and only if the social influence is strong enough.

In this case, and this is the second important result, the general analytical solution underlines the emergence of
critical endogenous thresholds in income and intrinsic attractiveness.
 The income threshold induces a segregation between those with an income above the threshold (whom we shall call the rich agents) and the others. These rich agents
 can freely choose their location, and they are the only ones to have access to locations above the threshold in intrinsic attractiveness. For agents below the income threshold, the housing possibilities are restricted to a subset of locations.
Applied to a monocentric city, this gives that: (i) people with a higher
 income live closer to the city center while the poorer people live in the periphery;
(ii) the rich agents are the only ones to live within a certain endogenous critical distance from the center, but there is no segregation amongst the rich agents within this central domain; (iii)
agents below the income threshold have access to locations at
distances greater than endogenous thresholds (the poorer the agent
the greater the critical distance); (iv) at any location outside the
center, except at the very periphery where only the poorest live,
there is always some social mixing. And this is the third important result: if the first of these
results, (i), is expected and in line with most of the literature,
the preservation of social diversity, on the one side amongst the richest
agents at the center, and on the other side in a large domain between
the center and the periphery,
 is particularly original and proved robust for a wide range of parameter values. 

The paper is organized as follows.
Section \ref{sec:model} presents the assumptions and the dynamics of the agent-based model.
 Section \ref{sec:param} reviews the key parameters of the model. Section
\ref{sec:analysis} deals with the mathematical analysis of the
equilibrium states. Section \ref{sec:segregation} describes the
conditions under which segregation occurs. Section \ref{sec:num}
presents numerical simulations for a simple case, that of a
monocentric city.  Section \ref{sec:data} considers a case allowing
for a better comparison between simulated and empirical data. The
conclusion follows.

\section{The agent-based model: assumptions and dynamics}
\label{sec:model} An agent-based model of residential location is
proposed. In this model, agents make their decision according to
both their income (characterized by an idiosyncratic willingness to
pay) and their individual evaluation of the level of attractiveness
of the different locations.

The assumptions of the model are presented below.

\subsection{Time, space and goods}
\label{sec:A0}

\begin{itemize}
 \item Time is discrete and indexed by $t$. The time
increment is $\delta t$ (in the numerical simulations in Sections
\ref{sec:num} and \ref{sec:data}, $\delta t=1$ is the numerical time
step;  in the mathematical analysis in Section \ref{sec:continuous},
the continuous limit $\delta t \rightarrow 0$ will be taken). The
horizon is infinite.
 \item We consider a 'city' defined as a discrete set $\Omega$ of locations $X$ uniformly
distributed on a bounded open set $\widetilde \Omega$ in $\mathbb{R}^2$.

The origin $O$ is taken as the geographical center of the city, and we denote by $D(X)$
the Euclidean distance from the center to a location $X$.
The total number of locations is $Card(\Omega)= L^2$, and the space is of linear size
(diameter) ${\cal D} = a L$, where $a$ gives the typical distance between two
neighboring locations (we therefore have $D(X) \leq a L/2$).
 \item In the city, there is a total number ${\cal N}$ of goods (housing for sale) with identical intrinsic
characteristics. The same number $N={\cal N}/L^2$ of dwellings is available 
at each location $X$ in $\Omega$.
\end{itemize}
\subsection{Agents}
\label{sec:A1}  At each period (given time $t$), there is a finite
number of agents in the economy, who can be in one of the three
following states: (1) buyer, (2) seller, (3) housed. We assume an
infinite ``reservoir'' of agents outside the city. Agents in the
reservoir are heterogeneous in their income - they are indiscernible
except for their income category.

From this reservoir, at each time step $t$ a constant number $\Gamma \delta t$ 
of randomly-chosen agents arrive on the market.
They are new
buyers who join the buyers who did not
buy a dwelling in the previous period. At the same time, housed agents become sellers at a homogeneous
rate $\alpha$. So the goods available for sale at a given location are those put on the market by
these agents plus, if any, those that have not yet sold.
Then matching occurs between buyers and sellers at each location (see below, Section \ref{sec:A6},
for the detailed rule).
Sellers who succeed in selling their good leave the market and return to the external reservoir. \\

Note that the total number of agents of each type - buyers, sellers and housed -
are dynamical variables since they depend on the success rate of the
previous period and on the inflow and outflow rates.
In what follows, the term "insiders" designates agents who are already housed (former successful buyers), while "outsiders" designates agents
looking for a flat.

\subsection{Demand prices}
\label{sec:A2}
Agents are characterized by their willingness to pay (hereafter WTP), which determines the maximum price the agent
is ready to pay for an asset. For simplicity, we consider a finite
number $K$ of WTP. Agents with the same willingness to pay are denoted by
$k$-agents, $k\in \{0,...,K-1\}$, and have WTP $P_k$. These WTP are ordered by
increasing values, $P_0 < P_1 <... <P_{K-1}$. When the agent is
acting as a buyer, his demand price is $P^d_k=P_k$. The agent's WTP also determines his behaviour as a seller, as explained below. 

Whenever needed, notably in Section \ref{sec:dist-crit}, the theoretical analysis will be done assuming an even distribution of the $P_k$ values between a minimum value, $P_{0}$, and a maximum value, $P_{0}+\Delta$:
\begin{equation}
 P_k=P_{0}\;+\; k\; \tfrac{\Delta}{K-1}, \;\;\;k=0,...,K-1,
\label{even}
\end{equation}
so that $\frac{\Delta}{K-1}$ is the constant increment between two consecutive prices.
In addition, throughout this paper we assume that these $K$ WTP values are uniformly distributed among the agents in the external reservoir. Hence, among the $\Gamma \delta t$ new incoming agents at any time $t$, a fraction $1/K$ has WTP equal to $P_k$, $k\in \{0,...,K-1\}$. These are not restrictive hypotheses, as shown in the Appendix, Section \ref{app:gen}.

\subsection{Attractiveness}
\label{sec:A3}
The attractiveness $A_k(X,t)$ of a location $X$ at time $t$, for a $k$-agent,
 depends on both (1) an \textit{intrinsic attractiveness}, $A^0(X)$, resulting from the location's intrinsic objective characteristics (e.g. local amenities), independent of the agent category,
  and (2) subjective characteristics
 which depend on the agent's social preferences, that is his preferences
concerning the social characteristics of the neighborhood. This attractiveness matters both when the agent is a seller (see \ref{sec:A4} 
below) and when he is a buyer (see \ref{sec:A6} 
below).

\subsubsection{The intrinsic attractiveness}
\label{sec:intrinsic}
The intrinsic attractiveness, $A^0(X)$,
idiosyncratic to the position considered, is here assumed to be time-independent. This means that we consider the time scale of transactions to be much shorter than that involved in the transformation of amenities, which we do not take into account.

\subsubsection{Attractiveness dynamics}
\label{sec:AttractDyn}
The attractiveness $A_k(X,t)$ is a \textit{subjective attractiveness}, whose value depends
 on the WTP of the agent looking at the location $X$. At each time step, new arrivals at the location $X$
modify the social composition at this location, and the attractiveness evolves accordingly.
Once the transactions between $t$ and $t+\delta t$ have occurred (as described below), the  attractiveness $A_k(X,t)$ of a location $X$ as seen by a
$k$-agent is updated according to:
\begin{eqnarray}
A_k(X,t+\delta t)= A_k(X,t)+\omega \delta t \, (A^0(X)-A_k(X,t)) + \; \delta t\; \Phi_k(X,t)
\label{eq:dAdtGal}
 \end{eqnarray}
where $\Phi_k(X,t)$ formalizes how the transactions at time $t$ affect the attractiveness according to the social preferences of the agents of WTP category $k$. We assume $\Phi_k(X,t)=0$ whenever no transaction occurred at time $t$:
the time evolution (\ref{eq:dAdtGal}) then implies that, when there is no transaction at a given location for a
certain amount of time, the attractiveness relaxes towards its
intrinsic value $A^0(X)$.
In the present work we assume that the attractiveness of a location for an agent increases when more agents with similar or higher incomes are housed at this location.
This is done by choosing $\Phi_k(X,t)$ proportional to the number of new buyers with WTP greater or equal to the agent's WTP. Hence we write:
 \begin{eqnarray}
A_k(X,t+\delta t)= A_k(X,t)+\omega \delta t \, (A^0(X)-A_k(X,t))) +
\epsilon \delta t\,  v_{k_{>}}(X,t)
\label{eq:At+dt}
 \end{eqnarray}
with
\begin{equation}
v_{k_{>}}(X,t)= \sum_{k'\geq k}v_{k'}(X,t)
\label{vk>}
\end{equation}
where $v_k(X,t)$ is the density of $k$-buyers in location $X$ who
complete a transaction at time $t$.
Here and in all that follows, what we mean by {\it density} is a number 

over the elementary surface area, $a^2$.

Note that
the level of attractiveness for a given WTP category depends on the intensity of the demand from agents of similar or higher category:
the higher the demand, the higher the level of attractiveness.

\subsection{Offer prices}
\label{sec:A4}
For each $k$-agent acting as a seller, his willingness to sell is determined by his WTP, $P_k$, and by the intensity of the demand, through the level of the mean attractiveness of
the location.
Specifically, the offer price $P^{o}_k(X)$ set by a $k$-agent selling a good at location $X$ is given by:
\begin{equation}
 P^{o}_k(X)=P^0+(1-\exp(- \xi \bar{A}(X,t)))\;P_k
\label{eq:bottom}
\end{equation}
where $P^0$ is the minimum price of an offer,
$\xi$ is a parameter, and $\bar{A}(X,t)$ is the average attractiveness of location $X$ at time $t$, that is:
\begin{equation}
\bar{A}(X,t)\equiv \frac{1}{K}\; \sum_{k=0}^{K-1}A_k(X,t).
\label{eq:Abar}
\end{equation}
For a good at location $X$
which has not
yet been exchanged, the offer price is given by:
\begin{equation}
\label{initial}
 P^{o}(X)=P^0+(1-\exp(-\xi\bar{A}(X,t) ))\;P^1
\end{equation}
Note that for $\xi$ small, the offer prices are all equal to $P^0$, with,
for $\xi$  small enough, $P^{o}_k(X)=P^0+ \xi \bar{A}(X,t)\,P_k $ and $P^{o}(X)=P^0+\xi \bar{A}(X,t)\,P^1$,
whereas for $\xi$ large, the offer prices are, respectively,
$P^{o}_k(X)= P^0+P_k$ and $P^{o}_k(X)= P^0+P^1$.

\subsection{Transaction prices.}
\label{sec:A5}
At location $X$, a transaction between a $k$-buyer (hence with demand price $P^d_{k}=P_k$) and a $k'$-seller with offer price
 $P^{o}_{k'}(X)$ given by (\ref{eq:bottom}),
can be completed
 if $P_{k} > P^{o}_{k'}(X)$.
If such a transaction occurs, the transaction price is assumed to be a linear combination of offer
and demand prices:
\begin{equation}
  P_{tr}=(1-\beta) P^{o}_{k'}(X)+ \beta P_{k}
  \label{up_price}
  \end{equation}
where $\beta$ is a constant coefficient.
Similarly, for a good exchanged for the first time, the transaction
price is $P_{tr}=(1-\beta) P^{o}(X)+ \beta P_{k}$,

with $P^{o}(X)$ given by (\ref{initial}).

\subsection{Market dynamics: the matching}
\label{sec:A6}
The market dynamics is then as follows, starting at time $t=0$ with an empty city and the attractiveness set to its intrinsic value. At each time step
(between times $t$ and $t+\delta t$):
\begin{enumerate}[leftmargin=0.5cm,itemindent=.5cm,labelwidth=\itemindent,labelsep=0cm,align=left]
\item  {\bf On the demand side.} For each $k \in \{0,...,K-1 \}$, the total number 
of $k$-buyers (outsiders) in the city is the sum of
 a constant number $\frac{\Gamma}{K} \delta t$ of new buyers (coming from the reservoir), and of
the $k$-buyers, if any, who are already present but have not yet bought a home. Each of these
agents decides to visit one particular location. The probability $\pi_k(X)$ for a $k$-buyer to visit a given
location $X$, depends on the attractiveness of the location:
\begin{equation}
\label{proba}
 \pi_k(X,t)=\frac{1-\exp(-\lambda A_k(X,t) )}{\sum_{X'\in \Omega}
1-\exp(-\lambda A_k(X',t) )}.
\end{equation}
For small $\lambda$ (that is for $\lambda \max_{k,X}A_k(X,t) <<1$), this is equivalent to:
\begin{equation}
\pi_k(X,t)=\frac{A_k(X,t)}{\sum_{X'\in \Omega}  A_k(X',t)}
 \label{lambda-small}
\end{equation}
These decisions, being made in parallel by all the outsiders, determine the demand at each location.

The above rule (\ref{proba}) can be seen as resulting from a search process where the agent requires a minimum acceptable level of quality. This latter is based on the relative attractiveness and on an additional component, corresponding to private information and/or local idiosyncratic characteristics of the good. This can be modeled as a random variable, so that in the end, the probability of selecting a location is an increasing function of $A_k(X,t)$.  The precise rule, Equ. (\ref{proba}), is not meant to assume some specific properties of the unknown random component, but is chosen for its simplicity and for the smoothness properties of the resulting search process.

\item {\bf On the offer side.}
At each location, the goods offered are (i) those which have not yet been sold, if any; (ii) those put on sale by the current owner at some previous time step and not yet sold, if any; (iii) goods newly put on the market by a fraction $\alpha$ of the remaining insiders.

\item {\bf The matching.}
We presented above the search process on the demand side. On the offer side, the search process is as follows. At each location, one of the sellers is selected by random draw (game against Nature); this seller then picks at random one of the potential buyers among those applying at this location and having a large enough WTP (see \ref{sec:A5} above), and the transaction occurs.
This is repeated until no offer or no buyer remains. Remember that sellers who succeed in selling their
good leave the market, returning to the external reservoir.

The matching rule is convenient for the mathematical analysis. In the numerical simulations, we also test an alternative matching rule: sellers are treated in order of increasing offer price, the buyers being randomly chosen as in the previous case.

\item {\bf Updating of the attractiveness}
Once the transactions have occurred, the attractiveness $\{A_k(X,t); k=0,...,K-1; X \in \Omega\}$ is updated as explained in \ref{sec:A3}
above, Equ. (\ref{eq:At+dt}).

\end{enumerate}

\section{Model parameters and key features}
\label{sec:param}
Let us now summarize the set of key parameters that characterize the model.

The model behaviour depends on the following set of control parameters:
(1) the incoming rate, that is the number $\Gamma$ of new agents entering the economy per unit of time; (2) the rate $\alpha$ at which agents leave the city; the parameters in the attractiveness dynamics, that is, (3) the time scale $1/\omega$ of relaxation towards the intrinsic value, and (4) the weight of the social influence, $\epsilon$; (5) the parameter $\xi$ which fixes the relevant scale of the attractiveness values in the offer prices (see Eq. (\ref{eq:bottom})). Finally, there are two structural control parameters: the intrinsic attractiveness, $A^0(X)$, and the social influence term, $\Phi_k$, in the dynamics of the attractiveness, Eq. (\ref{eq:dAdtGal}).

As we will see, the five control parameters only matter through three particular (dimensionless) effective parameters:\\

$\bullet$ The product

\begin{equation}
\widetilde \xi \equiv \xi A^0_{max}
\label{xitilde}
\end{equation}
where  $A^0_{max}$ is the maximum intrinsic attractiveness. 
The regime of interest is the one of not too large 
values of $\widetilde \xi$, for which

the offer prices are smoothly modulated by the attractiveness.

$\bullet$ The ratio of incoming to outgoing rates, $\frac{\Gamma}{\alpha}$. We are concerned with a parameter regime for which an equilibrium exists with no saturation (as explained 
in the next section). This requires
an equality between the mean inflow and outflow of agents, with a number of buyers that does not saturate the market.
One global necessary condition is that this ratio must be smaller than the number of goods per location, that is:
\begin{equation}
 \frac{\Gamma}{\alpha\, {\cal N} } = \frac{\gamma}{\alpha\,n}  < 1
\label{nonsat_cond}
\end{equation}
where
\begin{equation}
 \gamma\equiv \Gamma/{\cal D}^2
\label{defgamma}
\end{equation}
is the mean incoming rate per surface area, and
$n={\cal N}/{\cal D}^2=N/a^2$ is the density of goods (assumed to be uniform here).\footnote{Remark: for the multi-agent simulations it is convenient to define quantities per location,
whereas for the mathematical analysis with continuous limits, we consider densities.}

$\bullet$  An effective social influence parameter characterizing, in the building of the attractiveness, the strength of the social influence term compared to the intrinsic attractiveness:
\begin{equation}
 \eta \equiv \frac{\epsilon \gamma}{\omega \, \langle A^0 \rangle}
\label{eta}
\end{equation}
where the brackets $\langle . \rangle $ denote (here and in what follows) the spatial average over the city,
\begin{equation}
 \langle A^0 \rangle\equiv \frac{1}{L^2}\sum_{X\in \Omega}A^0(X).
\label{brackets}
\end{equation}
The origin of this parameter $\eta$ can be understood by looking at the dynamics (\ref{eq:dAdt}) and noting that
the maximum possible value of $v_{k_{>}}(X,t)$ is controlled by the incoming rate, $\Gamma$.

The analysis done in this paper
is for an arbitrary choice of $A^0$, and specified for particular cases, as previously explained.

What really matters in the present model are (1) the size of the domain over which $A^0$ is significantly non-zero (determining the 
part of the city on which transactions concentrate (see the Appendix, section \ref{sec:living})), and (2) the spatial heterogeneity in $A^0$ values. This shows up in the results of the following analysis of the stationary regime, where variables are found to depend on ratios such as
$ \frac{A^0(X)}{\langle A^0 \rangle}$ or $\frac{A^0_{max}}{\langle A^0 \rangle}$.\\

As already mentioned, different social preferences could be studied, by appropriate choice of the social term $\Phi_k(X,t)$. In particular, all the results in this paper straightforwardly generalize to the case where the agents weight higher categories differently:
\begin{equation}
\Phi_k(X,t)= \epsilon \sum_{k'\geq k}\, w_{k'-k} \,v_{k'}(X,t)
\label{wvk>}
\end{equation}
with weights $w_j \ge 0$ (see in particular the Appendix, Sections \ref{app:Proof_prop1}  and \ref{app:gen}).\\

Finally, we note that the other parameters are ``harmless'', that is their values  are not essential in controlling the model behaviour, provided they satisfy some global condition such as being 'small' or 'large': the time scale $\delta t$ (small in the mathematical analysis); the number of locations $L^2$, assumed to be large; the (large) city size ${\cal D}$ (or equivalently the small space scale
$a= {\cal D}/L$ between two nearby locations); the number $K$ of WTP categories (arbitrary, but it may be convenient to consider the large $K$ limit as providing a fine discretization of any continuous distribution of incomes); the set of WTP values (the $P_k$s) and the price parameters ($P^0, P^1, \beta$), which can be chosen so as to set price scales for comparison with any given empirical data; the total number ${\cal N}$ of goods in the city, with a number per location, $N={\cal N}/L^2$, assumed to be large in the mathematical analysis; the parameter $\lambda$ entering the probability to choose a location, Eq. (\ref{proba}), assumed to be  small (compared to the typical values of attractiveness), in which case it plays no role, except for providing smoothness properties with occasionally extremely high values of attractiveness.\\

Section \ref{sec:analysis} now mathematically characterizes the stationary regime that can be reached for a large range of parameter values.

\section{Theoretical analysis: dynamics and equilibrium}
\label{sec:analysis}

\subsection{Continuous time dynamics}
\label{sec:continuous}
For the mathematical analysis, the evolution of the above agent-based model is further formalized through partial differential
equations, taking the continuous time limit (that is $\delta t \rightarrow 0$).

In this continuous limit, the updating rule of the attractiveness $A_k(X,t)$ of a location $X$ seen by a $k$-agent, Equ.~(\ref{eq:At+dt}), gives:
\begin{equation}
 \partial_t A_k(X,t) =  \omega \,(A^0(X)-A_k(X,t)) + \epsilon \, v_{k_{>}}(X,t)
\label{eq:dAdt}
\end{equation}

Given the outgoing rate $\alpha$ (the fraction of agents leaving the inside per unit of time), and $v_k(X,t)$ being the density of newly-housed $k$-agents,
the density of insiders with willingness to pay $P_{k}$, $u_k(X,t)$, satisfies the following differential equation:
\begin{equation}
(1-\alpha)\partial_t u_k(X,t)=v_k(X,t)-\alpha\,u_k(X,t) \label{eq:dtu}
\end{equation}

The density of outsiders $ \rho_k(X,t)$ is written as:
\begin{equation}
\label{rho}
 \rho_k(X,t)=v_k(X,t)+\bar{v}_k(X,t),
 \end{equation}
with $\bar{v}_k(X,t)$ denoting the density of agents who did not
succeed in buying a good at location $X$ and time $t$.
Given the rules of the dynamics, the evolution of the density of outsiders $\rho_k(X,t)$ can then be written as:
\begin{equation}
\label{outsiders}
\partial_t \rho_k(X,t)  =  -\rho_k(X,t) \;+\;  \frac{\gamma}{K}\, L^2\,\pi_k(X,t) 
 \;+\;\pi_k(X,t)\,\sum_{X'\in \Omega}\bar{v}_k(X',t)
\end{equation}
In the above equation, $\frac{\Gamma}{K}\pi_k(X,t)$ is the rate at which new buyers visit $X$ at time $t$,

$\pi_k(X,t)$ being given by (\ref{proba}). The last term on the right-hand side of (\ref{outsiders}) is the fraction of agents, among those who did not buy a good at time $t$, who decided to search for a flat at location $X$.
Remark: note that the scaling gives a well-defined limit as $a \rightarrow 0$:
\begin{equation}
\label{outsiders_bis}
\partial_t \rho_k(X,t)  =  -\rho_k(X,t) \;+\;  \frac{\Gamma}{K}\,f_k(X,t)
 \;+\; f_k(X,t)\,\int  \bar{v}_k(X,t)\; d^2X
\end{equation}
where $f_k(X,t)=\pi_k(X,t)/a^2$ is the probability density.

\subsection{Equilibrium}
\label{sec:statio}
\subsubsection{Stationary state}
\label{sec:statio_gal}
We define  the \emph{equilibrium} as the stationary state (whenever it exists) of the
dynamics, i.e., the variables of the model - $\rho_k(X,t),
A_k(X,t)$,
$u_k(X,t)$ and $v_k(X,t)$ - become constant in time.
All the variables in the stationary state are
denoted with an asterisk $^*$.

Writing $\partial_t u_k=0,\; \partial_t A_k=0$ and $\partial_t \rho_k=0$,
in equations (\ref{eq:dtu}), (\ref{eq:dAdt}) and (\ref{outsiders}) respectively,
we obtain for the stationary state:
\begin{eqnarray}
v^*_k(X)&=&\alpha u^*_k(X),\label{v-stat} \\
A^*_k(X)&=&A^0(X)+\frac{\epsilon}{\omega} v^*_{k_{>}}(X) \label{A-stat} \\
\rho^*_k(X)&=&\pi^*_k(X)\;\sum_{X'\in \Omega}\rho^*_k(X').
\label{rhot-stat}
\end{eqnarray}
with
\begin{equation}
\pi_k^*(X)=\frac{1-\exp(-\lambda A_k^*(X) )}{\sum_{X'\in \Omega}
1-\exp(-\lambda A_k^*(X') )}
\end{equation}
or, in the small $\lambda$ limit (see (\ref{lambda-small}),
\begin{equation}
 \pi_k^*(X)=\frac{A_k^*(X)}{L^2 \langle A_k^* \rangle}
\label{pik*}
\end{equation}
with $\langle A_k^* \rangle= (1/L^2) \sum_{X'\in \Omega}  A_k^*(X')$.

Summing equation (\ref{outsiders}) on the whole space, we conclude that:
\begin{equation}
\frac{1}{L^2} \sum_{X\in \Omega}v^*_k(X)=\alpha \;\frac{1}{L^2} \sum_{X\in \Omega} u^*_k(X)
 =\frac{\gamma}{K}. 
\label{tot-stat}
\end{equation}
Consequently, in the stationary state, the mean 
density of $k$-housed agents on the whole space is equal to $\frac{\gamma}{\alpha K}$.
The total rate of $k$-transactions is $\frac{\gamma}{K}$. Thus, the total number of housed agents is the
same for all the $k$ values.\\
In the following, we restrict the analysis to equilibria in a non-saturated regime, as defined below.\\

\textbf{Definition: non-saturated equilibrium}.
{\it A non-saturated equilibrium is defined as an equilibrium where, for any given $k\in\{0,...,K-1\}$,
at any location $X \in \Omega$,  either the WTP of the $k$-agents are too low, so that none of them can buy a dwelling in this location, hence $v_k^*(X)=0$,  or the $k$-agents can afford to buy in this location, and in this case all $k$-demand is satisfied, that is:}
\begin{equation}
\bar{v}^*_k(X)=0.
\label{eq:nonsat_def}
\end{equation}

We denote by $\Omega_k$ the sub-set of locations $X$ where all the $k$-agents' demand is satisfied:
\begin{equation}
\Omega_k \equiv \{ X \in \Omega | \bar{v}^*_k(X)=0 \}
\label{omegak}
\end{equation}
$\Omega_k$ is the  set of locations (possibly empty, possibly identical to $\Omega$) 
where the $k$-agents are able to buy homes, and
the complement set, $\Omega \setminus \Omega_k$, is the set of locations where they cannot.\\

In what follows, the step-by-step analysis of the non-saturated equilibrium implies the unicity of such an equilibrium, whenever it exists. The existence conditions on the parameter values are the condition on the ratio of incoming to outgoing rates compared to the number of available goods (see Section \ref{sec:param}), the criterion (\ref{nonsat_cond}), and the conditions discussed in Section \ref{sec:frozen} concerning the offer and demand prices. 
As shown in Section
\ref{sec:num}, the above definition defines an equilibrium which is a
fairly good approximation of what is observed in the numerical
simulations. Results would be different on a market with rationing
(saturated case), as we will explain later.

\subsubsection{The conditions of segregation emergence: the WTP threshold} 
If the fraction of housed $k$-agents corresponds to the fraction of $k$-agents in the external reservoir and the ($k$-independent) intrinsic attractiveness,
it means that no segregation has emerged from the market dynamics.
Let us now characterize the subsets $\Omega_k$ of location possibilities for each category of $k$-agents.
The agents with the highest WTPs have access to any location, so that for them, $\Omega_k=\Omega$.
If a $k$-agent can afford a good at a given location, it is clear that $k'$-agents with $P_{k'}>P_k$ can as well:
we must have $\Omega_k \subseteq \Omega_{k+1} \subseteq \Omega$. Hence, in a stationary regime,
either $\Omega_k=\Omega$ for every $k$ (in which case there is no segregation), or there exists a critical value $\bar{k}$ such that for $k<\bar{k}$,
the demand of the $k$-agents is not satisfied on some of the locations, $\Omega_k \subset \Omega$ (and thus some socio-spatial segregation occurs). This means equivalently that there is a {\it WTP threshold}
 $P_c^* = P_{\bar{k}}$ so that only agents with WTP at least equal to this threshold can buy a good anywhere in the city.

Section \ref{sec:segregation} below sets out the existence conditions of such a WTP threshold $P_c^*$ associated with a critical category $\bar{k}$, together with
the characterization of the set $\Omega_k$ of locations
  where the $k-$agents with $k <\bar{k}$ can afford to buy a good.
We first determine the densities of housed agents in a non-saturated equilibrium whenever $\bar{k} \leq K-1$, distinguishing between categories above
the threshold, $k\geq \bar{k}$, and categories below the threshold,
$k <\bar{k}$.

\subsubsection{Housed densities} 
\label{sec:prop} 
For a non-saturated equilibrium, we determine the density of housed agents in terms of the intrinsic attractiveness and of the sets $\Omega_k$ introduced above.
We consider here the small $\lambda$ limit: 
we take (\ref{pik*})
for the probability of choosing a location.\\

\textbf{Proposition:}
In a  non-saturated equilibrium, whenever $P_c^* \leq P_{K-1}$, that is $\bar{k}\leq K-1 $:
\begin{itemize} \item for any WTP level $k\geq \bar{k}$, the density of housed $k$-agents does not depend on the WTP
level, but only on the intrinsic attractiveness
of the location, according to:
\begin{equation} \forall \textrm{$k\geq \bar{k}$ } \; \forall X \in \Omega, \; u^*_{k}(X)
=\frac{\gamma}{K \alpha} \frac{A^0(X)}{ \langle A^0 \rangle}
\label{dens}
\end{equation}
where $\langle A^0 \rangle$ is the average of $A^0$ as defined in (\ref{brackets}).
\item if $\bar{k}>0$ (that is $P_c^* > P_0$), for any WTP category $0 \leq k < \bar{k}$, that is with $P_k$ smaller than the WTP threshold $P_c^*$,
the density of housed $k$-agents
depends 
 on the intrinsic attractiveness of the location and on the set $\Omega_k\subset \Omega$ of locations in which the $k$-agents can afford to buy a good,
according to:
\begin{eqnarray}
 u^*_k(X) & = &\frac{\gamma}{K\alpha} \frac{A^0(X)}{\langle A^0 \rangle_k} \textrm{ if $X \in \Omega_k$ } \label{dens-cap}\\
& = & 0 \textrm{ otherwise}
\end{eqnarray}
with
\begin{equation}
 \langle A^0 \rangle_k \equiv \frac{1}{L^2} \sum_{X'\in \Omega_k} A^0(X').
\end{equation}
\end{itemize}

The proof of this Proposition, demonstrated by recurrence on $k$ starting from the highest WTP value, is given in the Appendix,
Section \ref{app:Proof_prop1}. 
This Proposition can also be shown to apply when the agents weight higher categories differently, as in Eq.~(\ref{wvk>}).

Eq.~(\ref{dens}) shows that for agents with high enough WTP ($k > \bar{k}$), the social influence parameter does not show up in the number of housed agents.
An intuitive explanation is that in a non-saturated regime, for ``rich'' agents not constrained by their income, the probability of settling at a given location depends on the attractiveness of
this location relative to that of all the other locations.

Eq. (\ref{dens-cap}) is an intuitive generalization of Eq. (\ref{dens}):
the normalization, that is the denominator in (\ref{dens-cap}), is given by the sum of $A^0$ over locations
where the considered $k$-agents can locate. For $k\geq \bar{k}$, by definition $\Omega_k=\Omega$, and (\ref{dens-cap}) reduces to (\ref{dens}).

\section{Thresholds and segregation}
\label{sec:segregation}
 Let us now determine the conditions for having the WTP thresholds within the range $[P_0, \;P_{K-1}]$, i.e., determine where people can be housed depending on their WTP category. In other words, let us characterize the sets $\Omega_k$.

The absence of segregation in the stationary regime would mean that, at any location, there are transactions involving any WTP category. If, on the contrary, at some locations the market dynamics make the distribution of offer prices higher than the distribution of demand prices, the poorest people cannot be housed, and thus some segregation occurs. In the worst case, the offer prices become so high that the intersection between the two distributions is null: this would mean that the system does not settle in a stationary state, but reaches a ``frozen'' state where housed agents cannot sell their good to any other agent. We determine here the conditions on the parameters that separate the regime with non-segregated equilibrium from the one with segregated equilibrium, and the regime with segregated equilibrium from the frozen state. This will also determine the thresholds in the intermediate parameter regime where segregation occurs.

\subsection{Stationarity criterion: non-profitable arbitrage}
\label{sec:criterion}
The existence of a stationary regime requires the absence of profitable arbitrage. Indeed, if at a given location, all the poorest inhabitants were selling their goods with a positive profit, this lowest category of income would eventually disappear from this location.
We thus write the stationarity condition, in other words, at each location, for each WTP level $k$ present in this location, 
the condition for the realization of transactions between two agents of the same category.
From the definition of the offer prices,
Eq. (\ref{eq:bottom}), we get the condition $P_k  \geq  P^0+(1-\exp(-\xi \bar{A}^*(X)))P_k$, which we can write as:
\begin{equation}
\label{minimum}
\xi \bar{A}^*(X) \; \leq \; \log \frac{P_k}{P^0}
\end{equation}
where we recall that $\bar{A}^*(X)=\frac{1}{K}\sum_k A_k^*(X)$.
As explained below, this condition enables us to identify the WTP category threshold $\bar{k}$.

\subsection{The frozen state}
\label{sec:frozen}
At the limit, nobody can be housed, and the attractiveness decreases towards the intrinsic value $A^0$.
 Then, a necessary condition for avoiding the frozen state is to have the criterion
(\ref{minimum})
satisfied for the highest income category, $k=K-1$, when the attractiveness is equal to the
maximum value that can be taken by the {\it intrinsic} attractiveness. This gives the condition:
\begin{equation}
\label{notfrozen}
\xi A^0_{max} \;< \; \log \frac{P_{K-1}}{P^0}.
\end{equation}

A frozen market could be linked to a market bubble. This sort of crisis could be attenuated by having adaptive agents with idiosyncratic $P^0$ and $\xi$ values: in order to sell their good, agents could lower at least one of these parameters, so that they would accept lower prices. Such a scenario would lead to an endogenous determination of some of the parameters. This clearly goes beyond the scope of this paper. We simply assume here that the $P^0$ and $\xi$ values are the same for every agent and within a range that allows to reach a stationary regime.

\subsection{The WTP Threshold}
\label{sec:wtp-thr}
We now write the conditions for the existence of a critical WTP level, $\bar{k} \in\{1,...,K-1\}$. Assuming that such a threshold exists, we get the expression of $\bar{A}^*(X)$ (see the Appendix, Section \ref{sec:Abar}):
\begin{equation}
\bar{A}^*(X)=A^0(X) \; \left(1 + \frac{\eta}{2}[\frac{K+1}{K}-\frac{(\bar{k}+1)\bar{k}}{K^2}]\right)
\end{equation}
where $\eta=\frac{\epsilon \gamma}{\omega \, \langle A^0 \rangle}$, as defined in (\ref{eta}).
The category threshold $\bar{k}$, and the associated WTP threshold $P_c^*=P_{\bar{k}}$, are obtained by writing the inequality (\ref{minimum}) for locations $X$ where $A^0$  reaches its maximum value, $ A^0_{max} $:
\begin{equation}
\label{crit-k}
\xi \,A^0_{max}  \, \left(1+ \frac{\eta}{2}[\frac{K+1}{K}-\frac{(k+1)k}{K^2}]\right)\;\leq  \;  \ln \frac{P_{k}}{P^0}
\end{equation}
and $\bar{k}$ is the smallest value of $k$ for which this inequality is true. 
\paragraph{Large $K$ limit}
For illustrative purposes, let us here assume that the WTP values are evenly distributed between
two extreme values, according to Eq.~(\ref{even}), that is
$P_k=P_{0}\;+\;\Delta\; \frac{k}{K-1}, \;k=0,...,K-1$ (see the Appendix, Section \ref{app:gen}, for the case of an arbitrary distribution).
In the large $K$ limit, writing that the threshold corresponds to the marginal case where the inequality becomes an equality, we obtain the WTP threshold $P_c^*$ as the solution of:

\begin{equation}
 \log \frac{P_c^*}{P^0} \,-\, \widetilde \xi \;=\;  \eta\;\frac{ \widetilde \xi }{2}\left(1 - \left(\frac{P_c^*-P_0}{\Delta}\right)^2 \right)
\label{eq:Pc}
\end{equation}
where $\widetilde \xi = \xi A^0_{max}$, as defined by Eq.~(\ref{xitilde}) Section \ref{sec:param}, is the effective parameter controlling the attractiveness dependency of the offer prices.

Since $\eta=\frac{\epsilon \gamma}{\omega \, \langle A^0 \rangle}$, as defined in (\ref{eta}),
note that the product $\eta \widetilde \xi$ on the right-hand side of (\ref{eq:Pc}) combines five key control parameters (see Section \ref{sec:param}): the inflow of agents $\gamma$, the social strength $\epsilon$, the relaxation time scale $1/\omega$, the offer prices parameter $\xi$, and a measure of the heterogeneity of intrinsic attractiveness, $\langle A^0 \rangle / A^0_{max}$.

\subsection{The emergence of segregation}
\label{sec:solutions}

Let us study the existence and solutions of Eq. (\ref{eq:Pc}). Analysis of the equation shows that there are three parameter regimes to consider, as detailed below.

$\bullet$ First, there is no solution for
$\widetilde \xi  > \log (P_{K-1}/P^0)$: this corresponds to the frozen state, where no transactions occur.

$\bullet$ Second, there exists a parameter domain where segregation occurs even in the absence of social influence. This is the case for $\log P_0/P^0 < \widetilde \xi \leq \log (P_{K-1}/P^0)$. In this case, a solution exists defining a critical threshold larger than $P_0$. Such a condition on the parameters is reached for large enough values of $\widetilde \xi$, that is if the offer prices are high, even if this is only due to the intrinsic attractiveness. For large values of $\widetilde \xi$,  Eq. (\ref{eq:bottom}) implies that the offer prices do not even depend on the attractiveness: $P^o_k=P^0+P_k$. 

\begin{figure}[!htbp]
\begin{center}
\includegraphics[width=9cm]{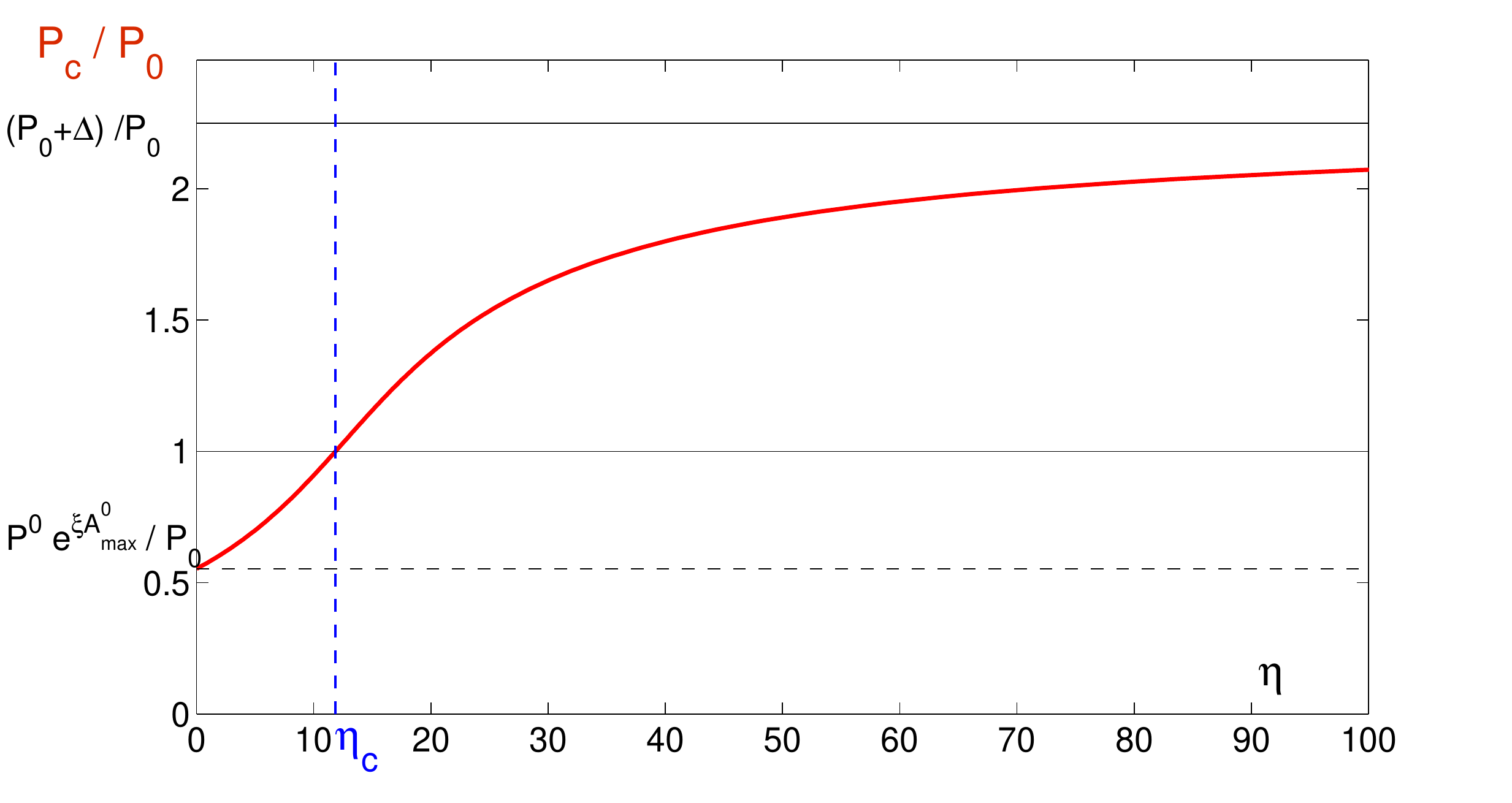}
\caption{Willingness to Pay Threshold $P_c^*$ (normalized by $P_0$) as function of the (effective) social strength $\eta$, illustrated for  $\Delta/P_0=1.25$ and $P^0 e^{\xi A^0_{max}}/P_0=0.55$. Segregation occurs for $\eta$  larger than the critical value $\eta_c$ given by condition $P_c/P_0=1$  (here $\eta_c=11.86$).} \label{fig:seuil}
\end{center}
\end{figure}

$\bullet$ The third case corresponds to
$\widetilde \xi < \log P_0/P^0$,
 illustrated in Fig.~\ref{fig:seuil}. As the figure shows, and as one can verify in Eq. (\ref{eq:Pc}), the threshold $P_c^*$ is an increasing function of $\eta$. This means that $P_c^*$ increases if the social factor $\epsilon$ increases, or if the inflow of agents increases; whereas $P_c^*$ decreases if the relaxation of the attractiveness towards the intrinsic value $A^0$ is faster (larger $\omega$), or if the domain of strongly attractive locations increases (larger $\langle A^0 \rangle$):  
and this is consistent with the economic intuition.

As $\eta$ goes from $0$ to $+\infty$, the critical threshold $P_c^*$ increases from
$P^0 e^{\widetilde \xi}$
to $P_0+\Delta$.
There is thus a critical value $\eta_c$ of $\eta$:
\begin{equation}
\eta_c=2\, ( \frac{1}{\widetilde \xi}\; \log \frac{P_0}{P^0} \,-\, 1)\; > 0,
\label{etac}
\end{equation}
below which $P_c^*$ is smaller than $P_0$, and thus every agent can locate at the center: there is no segregation. It is only when the social influence, $\eta>\eta_c$ is strong enough, that there exists a WTP threshold
that segregates the richest away from the others.
This is one of the most important qualitative results: the heterogeneity in $A^0$ values is not enough to lead to income segregation; in a non-saturated equilibrium, segregation emerges from
the strengthening of such heterogeneities, provided the strength of social influence is large enough.\\

For a given value of $\eta$, the condition for the emergence of segregation, $\eta > \eta_c$, is a condition on the parameters $P_0$, and $P^0 e^{\widetilde \xi}$, which characterize the smallest WTP value and the smallest offer value. 
Note that the threshold $\eta_c$ does not depend on WTP distribution, but only on the minimum value of WTP in the population, whereas the critical WTP value does depend on WTP distribution.

\subsection{Local and category thresholds}
\label{sec:local-th}
From the stationarity condition (\ref{minimum}), we can also deduce for each location $X$ the poorest housed category  $k_c(X)$ (that is the smallest value of $k$ such that the $k$-agents are present on $X$), and for each category $k<\bar{k}$ the maximum intrinsic attractiveness level, $A^0_{c, k}$, which defines the locations $X$ where the $k$-agents can be housed: $\Omega_k=\{X | A^0(X) \leq A^0_{c, k} \}$.

Consider $k<\bar{k}$.
Since $\langle A^0  \rangle_k \leq \langle A^0  \rangle$, from (\ref{minimum}) and (\ref{ak*}) we obtain the inequality:
\begin{equation}
\label{crit-k-X}
\xi \,A^0(X)  \; \leq  \;  \frac{\ln \frac{P_c^*(X)}{P^0}}{1+ \frac{\eta}{2}[\frac{K+1}{K}-\frac{(k_c(X)+1)k_c(X)}{K^2}]}
\end{equation}
with $P_c^*(X)=P_{k_c(X)}$.
This inequality can also be read as giving the maximum intrinsic attractiveness level $A^0_{c, k}$: 
\begin{equation}
\label{crit-A-X}
\xi \,A^0_{c, k}  \; \leq  \;  \frac{\ln \frac{P_k}{P^0}}{1+ \frac{\eta}{2}[\frac{K+1}{K}-\frac{(k+1)k}{K^2}]}
\end{equation}
Since the inequality (\ref{crit-k-X}) has been derived by replacing $\langle A^0  \rangle_k$ by $\langle A^0  \rangle$, we expect it to be strict. We get a lower bound on $k_c(X)$ (and a lower bound on $P_c^*(X)$, an upper bound on $A^0_{c, k}$) by writing that the category threshold $k_c(X)$ is the smallest value of $k$ for which this inequality is true - in  a continuous limit, it is the marginal value for which the inequality becomes an equality. As will be seen with the numerical simulations, these bounds are very good approximations.

\subsection{Case of a monocentric city: critical distances}
\label{sec:dist-crit}
Let us now apply the above results to the case of a monocentric city, with an intrinsic attractiveness monotonically decreasing from the center.  
Since the intrinsic attractiveness decreases from the center, the condition (\ref{crit-A-X})
means that for each $k$, the restricted set $\Omega_k$ is characterized by a
critical distance $d^*_c(k)$ from the center, beyond which the
$k-$agents' demand is positive.
For illustrative purposes, let us assume $A^0$ has a
two-dimensional Gaussian shape:
 \begin{equation}
 A^0(X)=A^0_{max} \exp\left(-\frac{D(X)^2}{R^2}\right)
\label{A0}
\end{equation}
where  $A^0_{max}$ is the maximum intrinsic attractiveness, $D(X)$ is the Euclidean distance from the center, 
and $R$ determines the distance from the center at which the intrinsic
attractiveness is still significantly non-zero. Then, the critical distance $d^*_c(k)$ 
satisfies:
\begin{equation}
 d^*_c(k)\geq R\sqrt{-\ln\left[\,\frac{\frac{1}{\xi A^0_{max}}\ln \frac{P_k}{P^0}}{1+ \frac{\eta }{2}[\frac{K+1}{K}-\frac{(k+1)k}{K^2}] } \,\right]}
\label{eq:dc}
\end{equation}

The above equation is valid whenever the argument of the logarithm is
smaller or equal to $1$, and otherwise $d^*_c(k)=0$.
This is the case for all $k$ levels above the WTP threshold, and the critical WTP level $\bar{k}$ is the smallest value of $k$ for which $d^*_c(k)=0$. The distance $d^*_c(\bar{k}-1)>0$ gives the limit of the central domain where only the richest can buy a dwelling.

\subsection{Summary}
\label{sec:summary}
Let us briefly summarize the main results of this section and comment on the consequences in terms of space organization (social mix or segregation).
  For the stationary state, we have shown that if and only if the social influence is strong enough, there emerges a WTP threshold which segregates the richest from the others.
The spatial distribution of agents with a willingness
 to pay higher than this critical value does not depend on
the  individual income level but only on the intrinsic
attractiveness of the locations. In other words, there is a social mix between agents with an
income above the WTP threshold. Agents with income below this threshold can only buy goods in a restricted set of locations, which depends both on the intrinsic attractiveness and on their income. In such locations, however, some social mix remains.

\section{Numerical simulations}
\label{sec:num}
We now simulate the agent-based model as presented in Section \ref{sec:model},
illustrating the theoretical results in the case of a pure monocentric
city.

\subsection{Simulation specifications}
\label{sec:num_param}

Following the rationale presented in Section \ref{sec:param}, 
the model parameters are chosen as follows. We consider a square lattice
of linear size $L=100$, hence with $L \times L= 10 000$ sites (locations). Each
location $X$ is characterized by its Cartesian coordinates $(x,y)$ in the
frame whose origin is the center of the square lattice.
Without loss of generality, time and space scales
are arbitrarily given by the time step $\delta t=1$ and
the lattice spacing $a=1$ respectively.
The intrinsic
attractiveness $A^0(X)$ is chosen to decrease with the distance from
the center, as in (\ref{A0}): we express it as a two-dimensional Gaussian function,
\begin{equation}
A^0(X)=A^0_{max}\exp(-\frac{D(X)^2}{R^2})
\label{A0bis}
\end{equation}
where $A^0_{max}$ is the maximum intrinsic attractiveness, $D(X)=(x^2+y^2)^{1/2}$ is the distance from the center, and $R$
determines the distance from the center at which the intrinsic
attractiveness is still significant. We set $A^0_{max}=1$ and $R=10$.

Most simulations are driven with $K=10$ categories,
 but we also present some results for $K=50$, which is large enough for comparison with the theoretical results obtained in the large $K$ limit.
The total number
$N$ of goods at each location is set equal to $200$. This is large enough to prevent the system
from saturating, i.e., to avoid having the number of assets for sale falling to $0$
on some sites.\\
The WTP values are set following Eq.~(\ref{even}), with $P_0=10^5$ and $\Delta=225000$. The offer price parameters are $P^0=9\,x10^4$, $P^1=2\,x10^5$ 
and $\xi=0.1$.\\
The other parameters are: the outgoing rate $\alpha=0.1$,
the inflow $\frac{\Gamma}{K}=1000$,
the social influence strength $\epsilon=0.022$, 
the attractiveness relaxation constant $\omega=\frac{1}{15}$,
the transaction price parameter $\beta=0.1$, and finally $\lambda=0.01$. \\
With these parameter values, the effective social strength parameter is $\eta =10.504$ for $K=10$,
and $\eta =52.521$ for $K=50$. 
In both cases, the critical value as given by Eq. (\ref{etac}) is  $\eta_c=0.107$, 
so that segregation is expected to occur. Finally, if one considers the distance unit as km and the price unit as euro, the results are of the same order of magnitude as what we observe for the Paris market (this is discussed in more detail in the next section). \\

The agent-based dynamics, detailed in Section \ref{sec:A6},

is briefly recalled here. At $t=0$, the simulated
city is empty, and the initial attractiveness of a location $X$
seen by every $k$-agent is set
to the intrinsic attractiveness: $A_k(X,t=0)=A^0(X)$.
At each time step, a randomly chosen fraction $\alpha$ of housed agents put their good on the market. On the demand side, a
number $\Gamma/K$ of new $k$-agents arrive from the external reservoir and join the
agents who failed to buy at the previous time step, if any. Every one of these agents chooses a location $X$
 where he would like to buy, the choice depending on the attractiveness of the locations
according to Eq.~(\ref{proba}).
In each location,
available offers are sold according to the chosen matching rule. Two sets of simulations are done, testing two versions of the matching rule (see Section \ref{sec:A6}):
\begin{itemize}
\item [(a)] the goods are sold in a random order, meaning that one buyer is picked at random for a randomly-chosen offer, then another buyer is picked a random for a second randomly-chosen offer, and so on.
\item [(b)] the goods are sold in order of increasing offer price, meaning a buyer is picked at random for the least expensive offer, then a second buyer is picked at random for the second least expensive offer, and so on.
\end{itemize}

Once all the transactions have been completed, the attractiveness is updated according to Eq.~(\ref{eq:At+dt}). The process is repeated until a stationary regime is reached.

In all cases, we find that the system reaches a
stationary state characterized by a constant number of housed agents on
the lattice. In what follows, we 
analyze these stationary
states. Focusing on the income and space distribution of housed agents, we compare the results of the numerical simulations with the results of the theoretical analysis when the matching rule is rule (\emph{a}). 

\subsection{Price distribution}
\label{sec:price-dist}
\begin{figure}[!htbp]
\begin{center}
\begin{minipage}[c]{.32\linewidth}
\includegraphics[width=4.5cm]{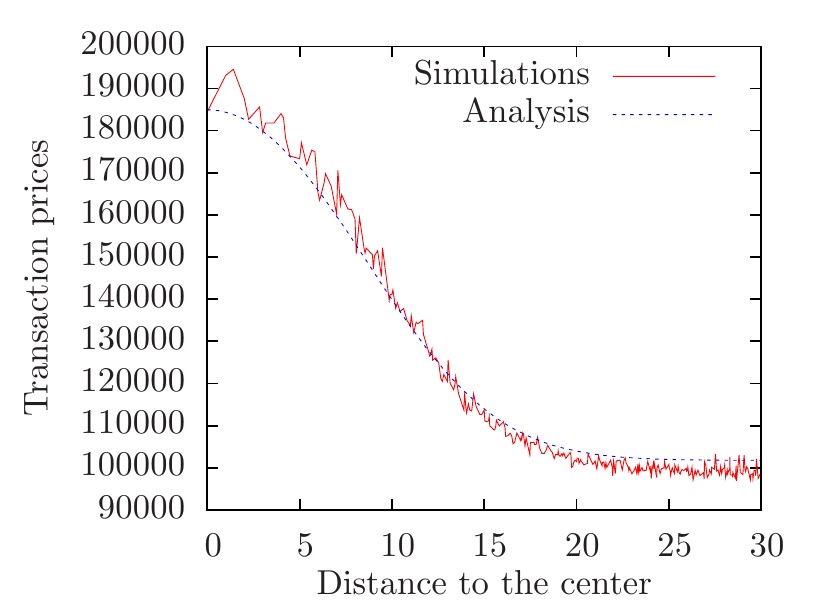}
 \end{minipage}
\begin{minipage}[c]{.32\linewidth} 
      \includegraphics[width=4.5cm]{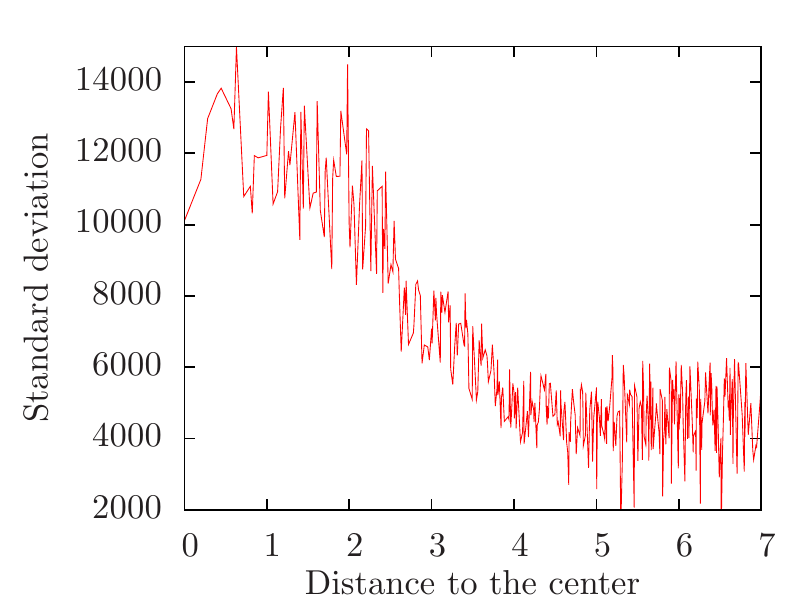} 
   \end{minipage}
\begin{minipage}[c]{.32\linewidth}
      \includegraphics[width=4cm]{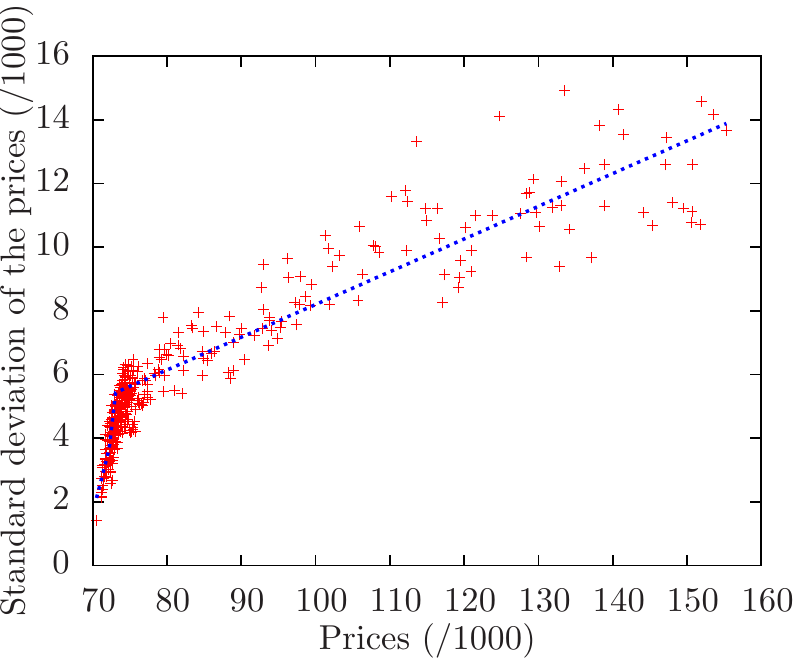} 
   \end{minipage}
\caption{Left: average prices (euros) as a function of the distance from the center, for $K$=50: the analytical result (solid curve) fits the numerical
simulations (dashed curve) well. Middle: standard deviation of transaction prices (euros) versus distance from the center;
 Right: standard deviation of transaction prices versus mean transaction prices (thousands of euros).
 The straight lines are linear regressions: (low prices) slope $0.9$,
 intercept $-62.000$ correlation coeff. $-1.0$; (large prices) slope $0.1$,
 intercept $-2.000$, correlation coeff. $-0.97$.}
\label{SD-sim}
\end{center}
\end{figure}

Once a stationary regime has been reached, we measure the mean transaction prices and the associated
variances. For the price distributions, we find no noticeable differences when rule (a) or rule (b) is selected, and Fig.~\ref{SD-sim} presents the results for rule (a).
As expected from the theoretical results, transaction prices decrease from the center to the periphery (Fig.~\ref{SD-sim}, left). Looking at the standard deviations (Fig.~\ref{SD-sim}, middle), we can distinguish two parts: in the city center,
  a large distribution of the prices shows an active dynamics,
   whereas in the near peripheral area, we observe a trend towards price
    homogenization.
  The interesting fact that prices present a larger variance in the central area than at the
 periphery, i.e., the variance is higher for higher mean prices (Fig.~\ref{SD-sim}, right), is consistent with the theoretical analysis in Section \ref{sec:statio}:  in locations where only the richest can afford to buy a good,  
there is no segregation amongst these rich agents. Hence there is a large variation of transaction prices in these locations.

\subsection{Socio-spatial segregation and income mix}
\label{sec:revenue}

\begin{figure}[!htbp]
\begin{center}
   \begin{minipage}[c]{.46\linewidth}
      \includegraphics[width=6cm]{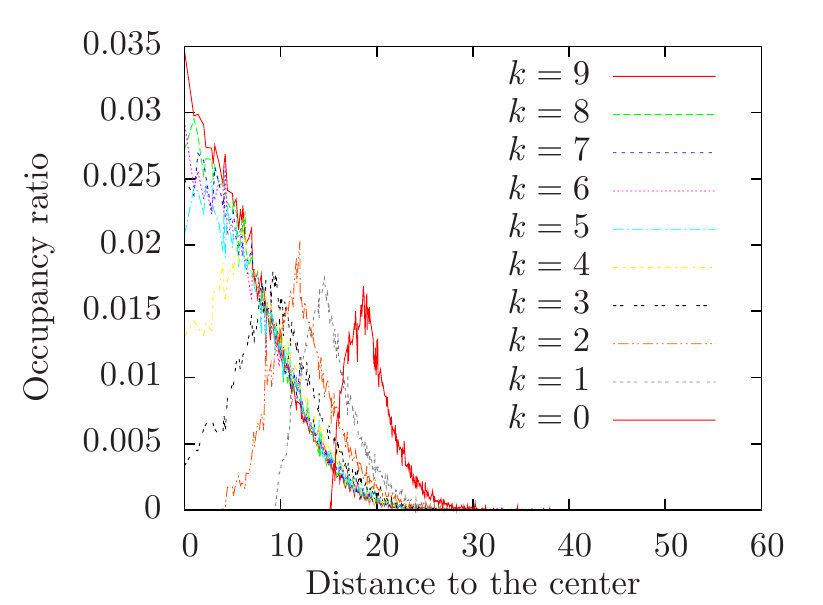}
   \end{minipage} \hfill
   \begin{minipage}[c]{.46\linewidth}
     \includegraphics[width=6cm]{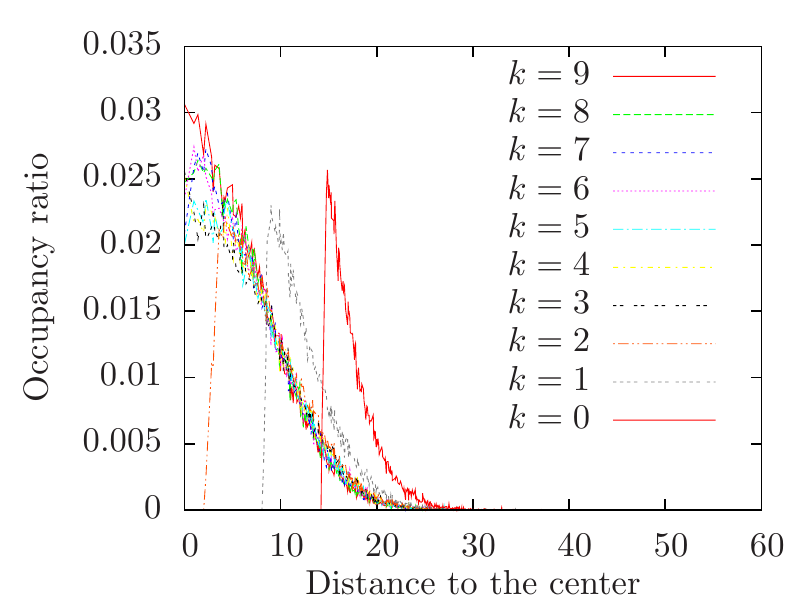}
   \end{minipage}
\caption{Occupancy ratio per WTP versus the distance from the
center in the stationary state. $K=10$ income categories are present on the
lattice. The distance from the center is in arbitrary units. The offers to be sold are randomly chosen (case (a), right panel), or in order of increasing prices
(case (b), left panel).}
\label{occup}
\end{center}
\end{figure}

Fig.~\ref{occup} shows the occupancy ratio for the
different WTP in the stationary state with respect to the
distance from the center. People
are located inside the zone where the intrinsic attractiveness
is significantly non-zero, in keeping with the analysis of Section \ref{sec:living}.

\begin{figure}[!htbp]
\begin{center}
\includegraphics[width=6cm] {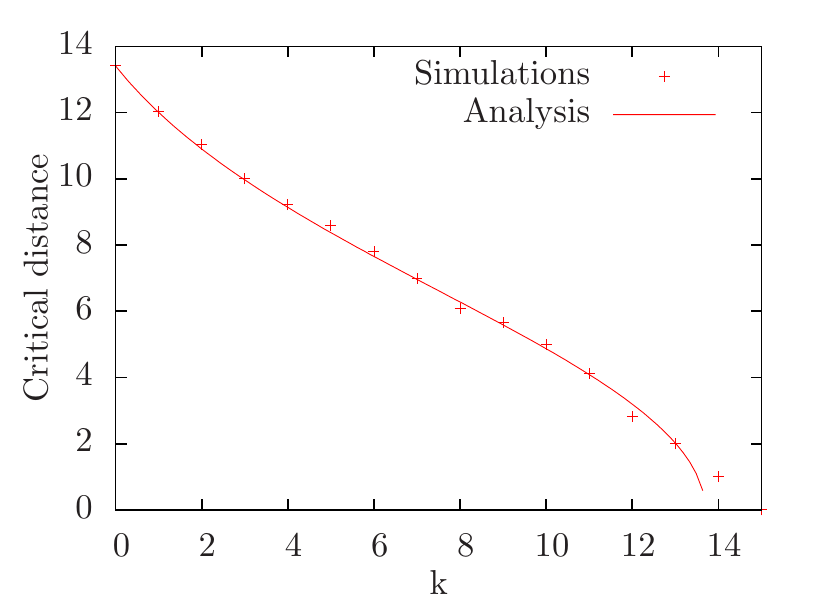}
\caption{Critical distance with respect to the WTP category $k$
($K=50$): the points are obtained from the simulations, the solid
curve is the analytical bound (the right-hand side of (\ref{eq:dc})).} 
\label{crit-dist}
\end{center}
\end{figure}

Figure~\ref{occup} compares the distribution of housed people with two different search rules: in case (a), the density of $k$-agents excluded from the center jumps discontinuously to a non-zero value at a critical distance from the center, whereas
 in case (b), the transition is smoother. However, the overall distribution is the same, confirming the theoretical predictions obtained for case (a): two zones of segregation, the rich and the poor, and in between an intermediary zone with some  social mix.

\begin{figure}[!htbp]
\begin{center}
   \begin{minipage}[c]{.46\linewidth}
      \includegraphics[width=5.5cm]{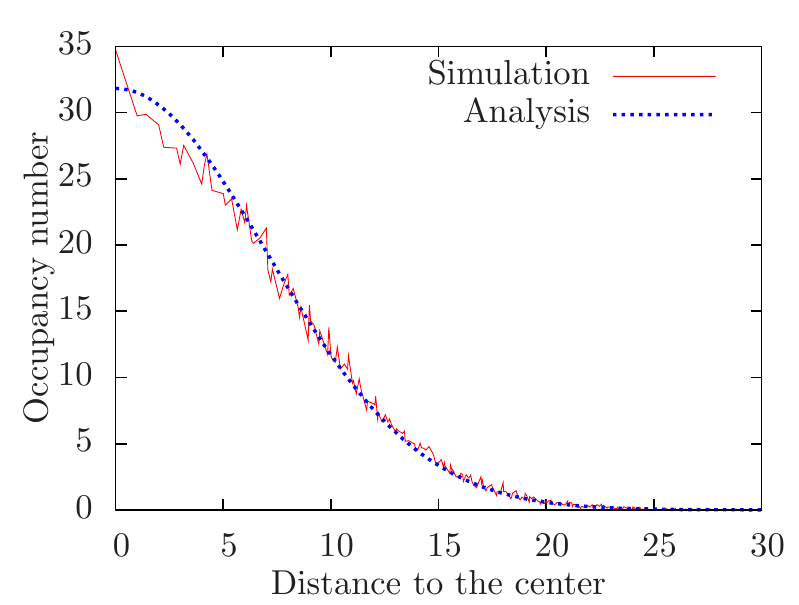}
   \end{minipage} \hfill
   \begin{minipage}[c]{.46\linewidth}
     \includegraphics[width=5.5cm]{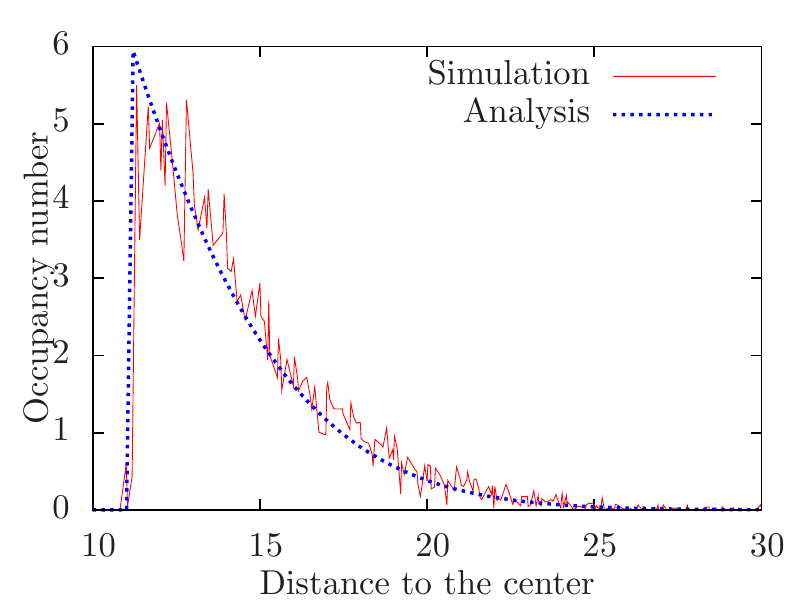}
   \end{minipage}
   \caption{Mean number of housed agents with respect to the distance from the center ($K=10$).
Left: for rich agents (income level $k=9$). Right:  for poor people (income level $k=2$).
The solid curves correspond to the simulations,
 while the dotted ones are from the theoretical analysis: Left, Equ. (\ref{dens}), right: Eq. (\ref{dens-cap}), with $d^*_c(k)$ replaced by its lower bound, the right-hand side of (\ref{eq:dc}).}
\label{analyse-sim-hbc}
\end{center}
\end{figure}

The critical WTP value $\bar k$ is found to be around $15$, and the agents with a lower WTP are distinguished by their income: the lower the WTP, the higher the distance from the center.
Fig.~\ref{crit-dist} shows how the critical distance increases as the WTP level decreases.
Fig.~\ref{analyse-sim-hbc} 
represents the number of agents housed in a given area, on the left for one category of rich agents, and on the right for one category of poor agents. Both figures exhibit a good correlation between the analytical results and the numerical simulations.\\

To summarize, when the offers are randomly selected - rule (a) -, 
the numerical results fit the non-saturated equilibrium defined in the analytical part: at some distance specific to the agent's WTP, one has a sharp transition from no housing ($v(X)=0$) to complete housing (all potential buyers
become housed, $\bar{v}(X)=0$).
In the case where the less expensive assets are sold first - rule (b) -, there exists a small domain of locations where both $v(X)$ and $\bar{v}(X)$ are non-null.
We recall, however, that for the distribution of transaction prices, the two rules give the same results.

\subsection{Social mix index} \label{social_mix}
Now we will use quantitative measures of segregation to estimate the level of social diversity.
There is a large literature on such measures: here we consider two measures particularly adapted to our study.
First, we introduce a multigroup social mix index
 derived from the mean relative deviation index proposed by \cite{ReardonFirebaugh2002}:
\begin{equation}
 ID(X)=\sum_{k=0}^{K-1} |\nu_k(X)-\frac{1}{K}|
\label{eq:ID}
\end{equation}
where $\nu_k(X)$
is the relative density of $k$-agents:
\begin{equation}
 \nu_k(X) \equiv \frac{u_k(X)}{\sum_{k'=0}^{K-1} u_{k'}(X)}
\end{equation}
This gives the
difference between the uniform distribution and the observed
distribution: the larger
this index, the greater the segregation
at this location.

Second, we use the entropy (or Shannon information) associated with the income distribution at each location as a measure of segregation.
\begin{equation}
 H(X)= - \sum_{k=0}^{K-1} \nu_k(X) \;\log \nu_k(X).
 \label{eq:H}
\end{equation}
The larger the entropy, the weaker the segregation (with a maximum possible value of $H=\log K$ when all categories are equally present, and with $H=0$ as the minimum value when a single category is found at the considered location). 
The entropy has unique mathematical properties that are appropriate for a measure of heterogeneity\footnote{In particular, it allows the coherent study of segregation at different scales \emph{e.g.} if one wants to see how the measure changes when increasing the number of income categories.} - for a discussion in the economics context, see \cite{Theil67}.

In contrast to most of the literature, we do not consider averages of the indices  over all the spatial locations (the social mix index $ID(X)$ and the entropy $H(X)$ are defined for each location $X$). To quantify the segregation as a function of the distance from the center, we consider the averages of the indices $ID$ and $H$ over locations situated at a same distance from the center.
\begin{figure}[!htbp]
\begin{center}
   \begin{minipage}[c]{.45\linewidth}
      \includegraphics[width=6cm] {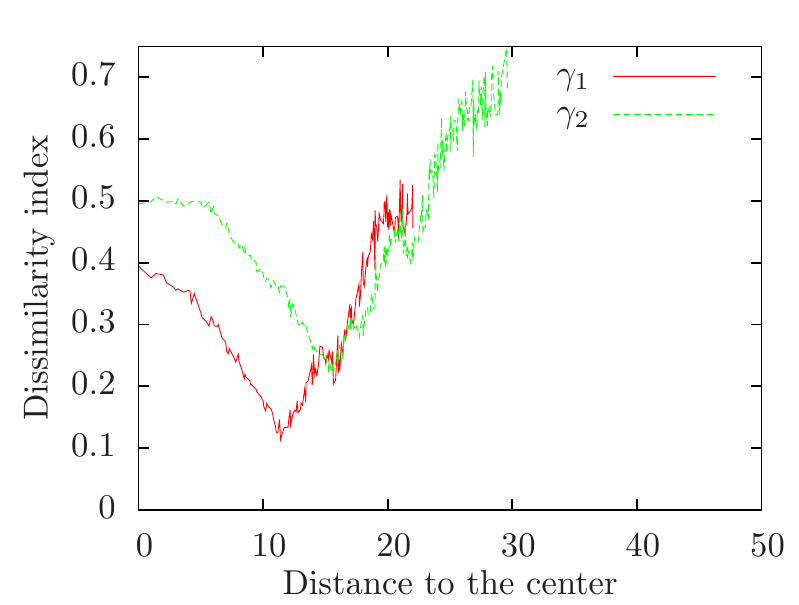}
   \end{minipage} \hfill
   \begin{minipage}[c]{.45\linewidth}
      \includegraphics[width=6cm] {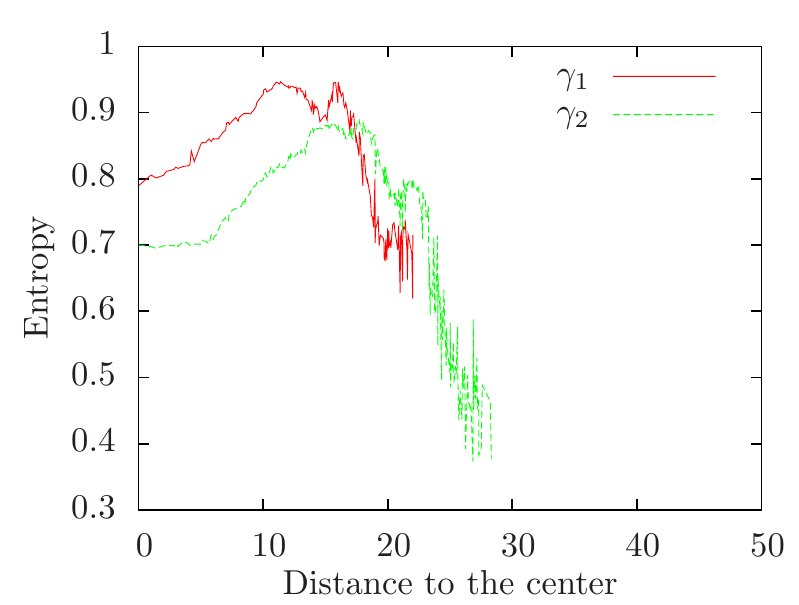}
   \end{minipage}
\caption{Social mix index $ID$ (left) and  Entropy $H(X)$ (right), 
as functions of the distance f,om the center, with $K=10$, for two rates of newcomers  $\gamma_1=1$ (that is $\Gamma/K=1000$) and $\gamma_2=3$. Other parameters as for the other simulations.}
\label{ID}
\end{center}
\end{figure}
Figure \ref{ID} shows theses average values, comparing two different rates of newcomers, $\gamma_1 < \gamma_2$.

The plots quantify what we inferred from Fig.~\ref{occup}: there is a zone near the city center 
with a moderate value of each index, a peripheral zone with a high index (low entropy), and
an intermediate zone with a weak index (high entropy). This confirms
the presence of an intermediate area of social mix. We also see that when we increase the rate of newcomers and hence the local demand, the index (resp. entropy) is on the whole
higher (resp. lower), meaning there is a general decline in the social mix.
The domain of social mix becomes smaller, and income segregation is thus amplified. 

\subsection{Saturated regime}
\label{sec:saturated}
\begin{figure}[!htbp]
\begin{center}
\includegraphics[width=6cm] {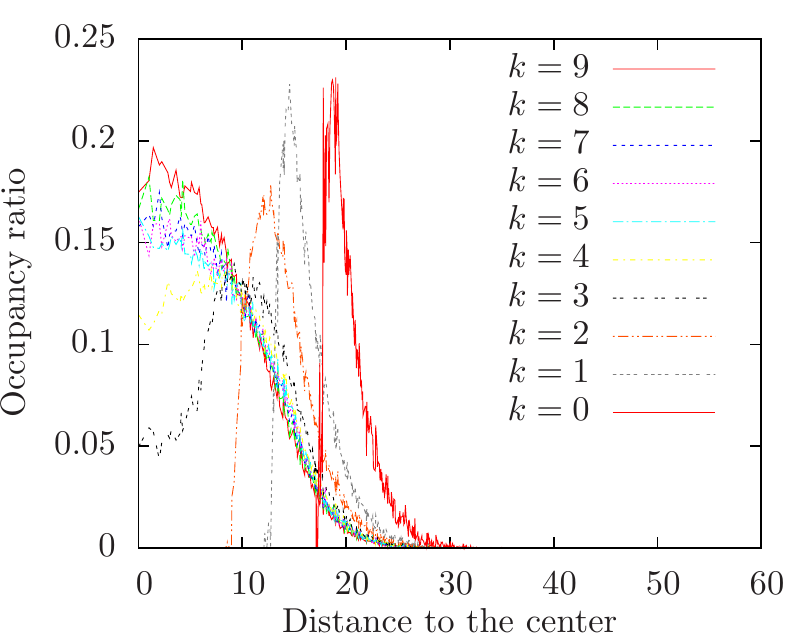}
\caption{Saturated stationary regime: occupancy ratio per WTP versus the distance from the center (in arbitrary units). 
 The parameters of the
simulations are the same as in Fig.~\ref{occup} except for $\Gamma/K=2000$.}

\label{occup-sat}
\end{center}
\end{figure}

Increasing the incoming rate further eventually
 leads to a saturated state, i.e., with a situation of excess demand at some locations.
An example is illustrated in Fig.~\ref{occup-sat}. The
parameters of the simulations are the same as before, except for
the rate of newcomers which is higher.
Comparing with Fig.~\ref{occup}, we see that saturation increases the
critical distances for the low WTP agents: the poorer agents move
further away from the center. 

The analytical expressions derived previously do not hold
in this saturated case, for which the mathematical analysis 
requires a specific study beyond the scope of the present paper.
Nevertheless, the simulations indicate that the
distribution of the population exhibits similar features to those obtained in the non-saturated case, with increased income segregation.

\section{Application to the Paris housing market}
\label{sec:data}

This section proposes a first comparison between the results of the model -
analytically demonstrated and validated by simulations - and empirical observations. The aim here is to see how far the price distribution in
Paris can be explained by the phenomena described above.

\subsection{The empirical database}
\label{sec:Paris-data}
The main reliable information source on real-estate prices in Paris is the
B.I.E.N. database, managed by the "Chambre des Notaires de Paris",
which records real-estate transactions for Paris and the Ile-de-France region.
This database covers all categories of real estate, indicating, for each transaction, more than 100 different elements of information drawn from the associated legal documents.
 For our study, we extract the information we need, concerning the locations and transaction
prices for flats. Since the data base contains information
from $1990$ until $2004$, the long term dynamics may be studied: this
will be the subject of future works, requiring additional features in
the model, notably the introduction of a slow dynamics for the
intrinsic attractiveness. In the present work we restrict the
analysis to one particular year, namely $1994$, for which the number
of registered transactions is large enough. Moreover, this is an important year because it is situated before the fall in prices which affected Paris from 1995 to 2000.

During the year $1994$, about $13 000$
transactions were recorded in Paris. The average price of a flat was $143 300$ euros, the standard deviation was around $90 000$ euros and the
distribution was not normal (the normalized kurtosis
coefficient is strictly positive). Since the database does
not contain the incomes of the buyers, we use the transaction
prices as a proxy for income distributions.

\begin{figure}[!htbp]
\begin{center}
\includegraphics[width=9cm]{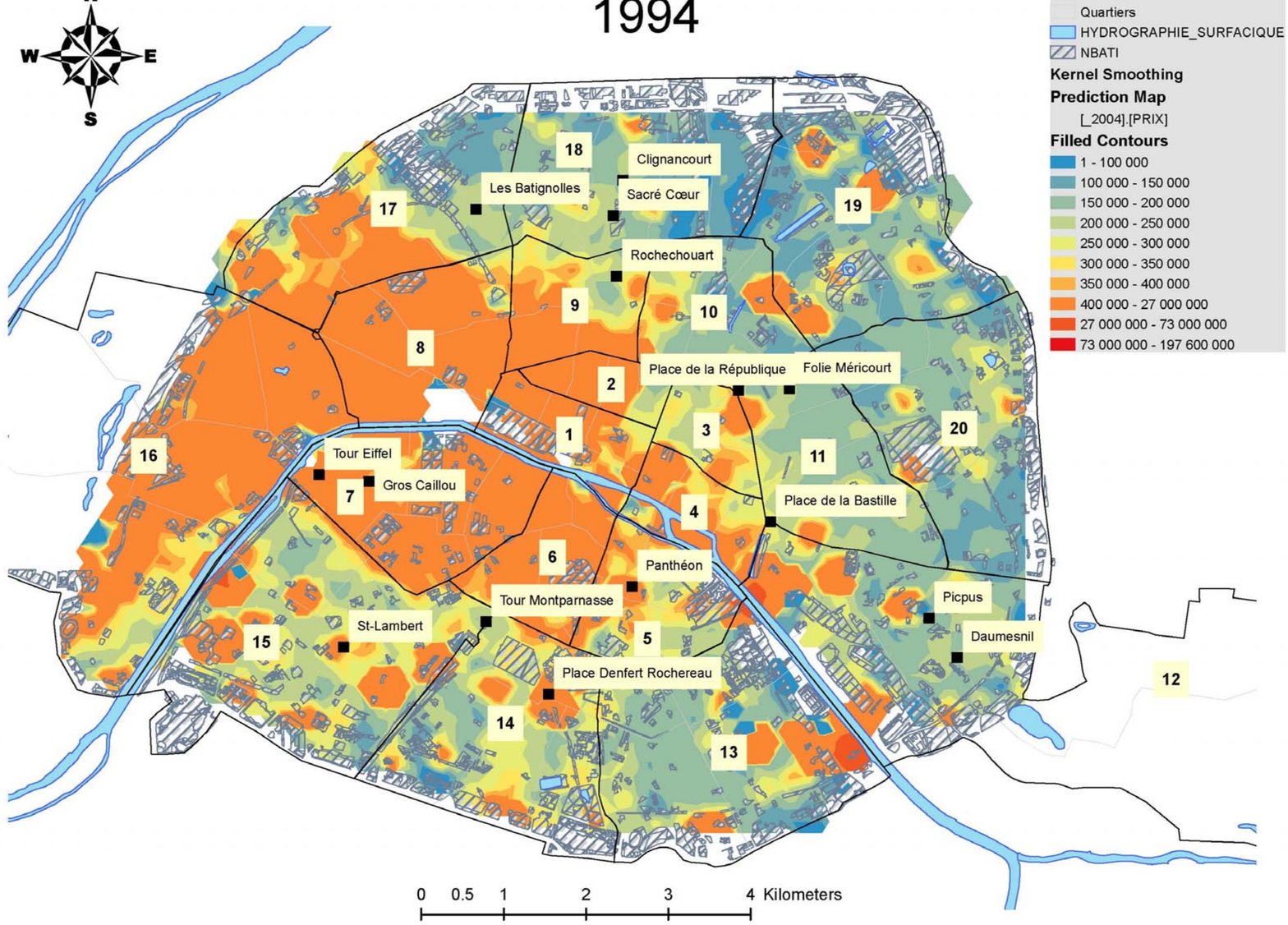}
\caption{Transaction prices in Paris in the year $1994$.
}
\label{data-paris}
 \end{center}
\end{figure}

\begin{figure}[!htbp]
\begin{center}
   \includegraphics[width=7cm]{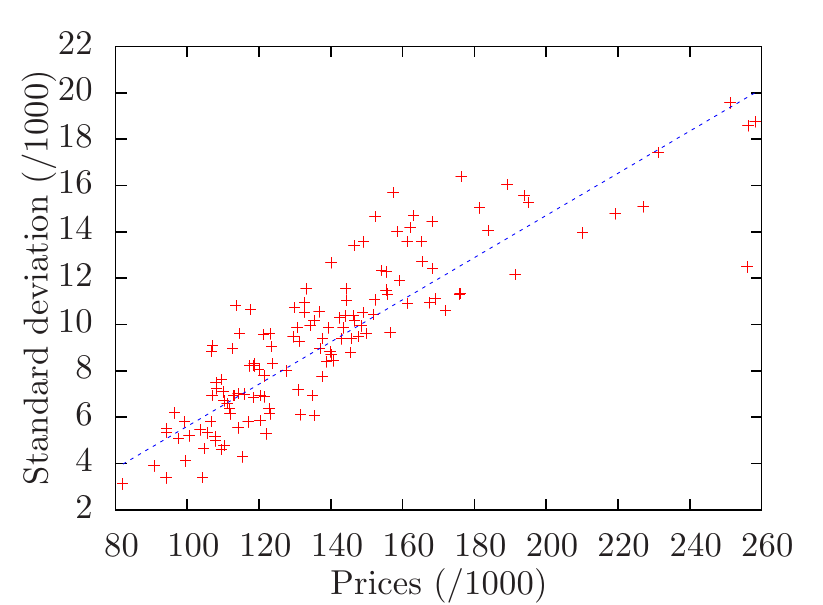}
   \caption{Paris data. Standard deviation versus transaction
   prices in thousands of euros. The straight line is the corresponding linear regression:
   slope $0.091$, intercept $-3.482$, correlation coeff. $-0.972$.}
\label{prix-var-data}
 \end{center}
\end{figure}

Fig.~\ref{data-paris} and Fig. \ref{prix-var-data} give some idea of the spatial distribution of prices in Paris. Paris is not a very large city, and its geometrical form ('like a potato')
endows the geographical center, identified as the historical center {\it Notre-Dame de Paris}, with particular importance. Looking at the map of the prices in Fig.~\ref{data-paris}, a
first obvious fact is that prices are high in the center,
with a marked trend of decreasing prices as one moves away from the center. Hence, the
monocentric model could be seen as a first-order approximation of the Paris market.
However, there are obvious deviations from this general trend. In particular, the "Rive
Gauche" effect is important: at a similar distance from the centre, prices are higher
on the left bank than on the right. In addition, prices are high in the
16th \textit{arrondissement} on the very outskirts of Paris, creating a ``hot-spot'' of high prices far from the center. This map is obtained by averaging transaction prices over areas of 500 by 500 meters.

In Fig. \ref{prix-var-data}, standard deviations are plotted with respect
to the corresponding average prices used in Fig.~\ref{data-paris}, for each of the areas that contain enough transactions for the measure to be relevant. As expected, the standard deviation grows with the prices, suggesting a higher diversity when prices are high. However, it should be noted that the standard deviation is an {\it affine} function of the average price.\footnote{This observation would have been less evident had we normalize the standard deviation by the mean.} This behaviour is remarkably similar to that obtained in the model, both qualitatively and quantitatively: compare
the plot in Fig.~\ref{prix-var-data} with Fig.~\ref{SD-sim}, right (case of the modeled monocentric city), and the linear regression slopes in both cases.

\subsection{Data-driven modeling}
\label{sec:Paris-model}
As explained above, a simple monocentric model would not allow to fit the price distribution in Paris.
Here we simulate the Paris prices by considering a more specific model of the city, which combines a general preference for the center together with preferences for some local particularities.

\begin{figure}[!htbp]
\includegraphics[width=7cm]{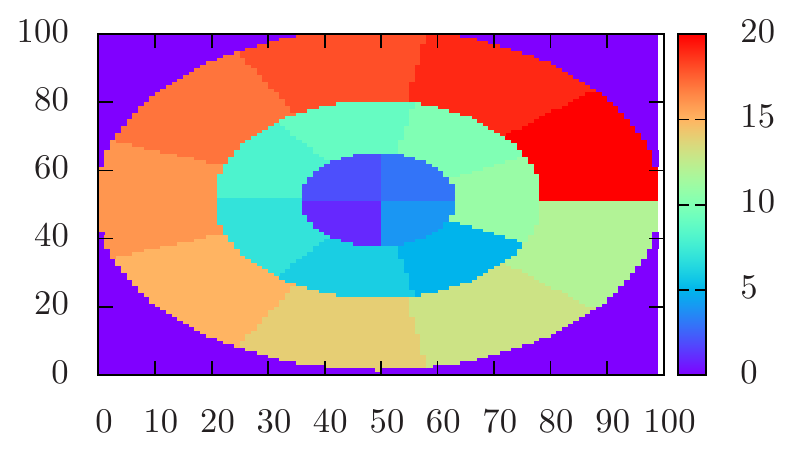}\hfill
\includegraphics[width=7cm]{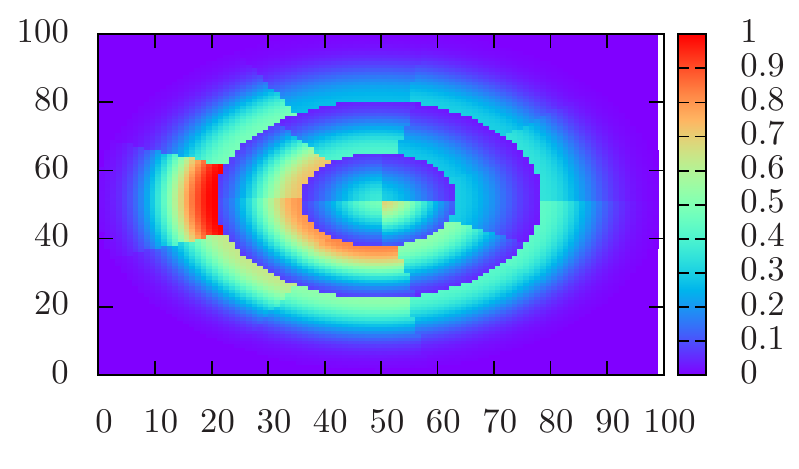}
\caption{On the left: stylized map of Paris. On the right: map of the intrinsic attractiveness. $P_{ref}^1=151002;$
   $P_{ref}^2=119440;$
   $P_{ref}^3=149606;$
   $P_{ref}^4=142332;$
   $P_{ref}^5=157623;$
   $P_{ref}^6=188535;$
   $P_{ref}^7=190785;$
   $P_{ref}^8=203960;$
   $P_{ref}^9=146715;$
   $P_{ref}^{10}=119515;$
   $P_{ref}^{11}=121604;$
   $P_{ref}^{12}=131973;$
   $P_{ref}^{13}=134051;$
   $P_{ref}^{14}=152234;$
   $P_{ref}^{15}=162560;$
   $P_{ref}^{16}=225985;$
   $P_{ref}^{17}=150991;$
   $P_{ref}^{18}=104776;$
   $P_{ref}^{19}=115073;$
   $P_{ref}^{20}=110912;$ $R_1=13.6;
    R_2=28.9;
    R_3=50$.}
\label{map-model}
\end{figure}

First, instead of using a square lattice in the simulations, we use a stylized map of Paris. It consists of three concentric zones of radius $R_1$,$R_2$,$R_3$, each zone
with $4$,$7$ and $9$ areas respectively representing the \textit{arrondissements}, and the ratios of the simulated areas fitting the real ratios well. The stylized map of Paris is presented on the left of Fig.~\ref{map-model}.
Second, we assign to each one of these ``\textit{arrondissements}'' a specific intrinsic attractiveness, decreasing with distance from the city center.

For each \textit{arrondissement} in the model, the intrinsic attractiveness parameters are scaled according
to the average transaction price of the corresponding (real) \textit{arrondissement} in the year $1994$.
More precisely, the maximum attractiveness is taken
as the ratio of the squared mean transaction price in the \textit{arrondissement} studied to the squared maximum mean transaction price observed among all the
 \textit{arrondissements}.
Finally, the intrinsic attractiveness $A^{0,a}$ of each \textit{arrondissement} $a=1,...,20$ is given by:
\begin{equation}
 A^{0,a}(X)=\left( \frac{P_{ref}^a}{P_{ref}^{max}} \right)^2 \; exp\left(-\, \frac{(R^a-D(X))^2}{R^2}\right)
\label{Att-data}
\end{equation}
where $D(X)$ is the distance from the center, $R^a$ the shortest
distance from the \textit{arrondissement} to the center (which can take one of the three values, $R_1, R_2, R_3$), and $P_{ref}^a$ the
mean transaction price of the \textit{arrondissement} calculated for the
year $1994$.
The value of $R$ is chosen to be of the same order of magnitude as the mean distance to the center of the lattice. The resulting map of the intrinsic attractiveness is presented in the right-hand panel of Fig.~\ref{map-model}.

Note that firstly, the above form and parametrization of the attractiveness gives a stronger attractiveness to the more expensive \textit{arrondissements}, even if they are far from the center of the city, and secondly, the chosen calibration only provides the overall price trend of the \textit{arrondissements} - it
does not determine how the prices vary within a given \textit{arrondissement}.

For the year $1994$, we order the transaction prices and divide the
price domain into intervals such that each one includes about $1000$ transaction prices. We postulate
that prices belonging to a same price interval correspond to transactions made by
individuals belonging to a same income class. In the model, each of these intervals is thus associated with one WTP category. This gives $K=13$ WTP levels in the model.

According to data published by the INSEE (the French National Institute of Statistics and Economic Studies), there are about
$375 000$ home-owners in Paris (we only consider main residences),
and about $13 000$ transactions in
one year. We therefore deduce a rate of house-moving of about $3.5\%$, hence $\alpha=0.035$.
Lastly, we take $\frac{\Gamma}{K}=1000$,
$\epsilon=0.18$, 
$R=10$, $\lambda=0.005$, $\xi=0.05$, $\Delta=225000$,
$\beta=0.1$, $P^0=90000$, and $P_0=100000$. Assuming the price unit to be the euro,
this set of parameter values allows us to obtain transaction prices of the same order as the empirical prices.

\paragraph{Arrondissement-specific segregation}
When applied to the present case, the general theoretical results predict the emergence of segregation with critical distances specific to each \textit{arrondissement}. The space is here decomposed into the union of
 $m=20$ disjoint sub-spaces (\textit{arrondissements}) $\Omega^a, a=1,...,m$,
 such that, on each of these sub-spaces, the intrinsic attractiveness is
  continuous and monotonically decreasing with the distance from the center of the city.

Each $\Omega_k$ (the subset of locations to which $k$-agents have access) is then
the union of sub-sets $\Omega_k^a$,  where the agents
 have a sufficiently high WTP to exchange,
$\Omega_k = \bigcup_{a=1}^m \Omega_k^a$
with
\begin{equation}
       \Omega_k^a \equiv \{X\in \Omega^a  | D(X)\geq d^a_c(k) \}
\label{eq:omega}
      \end{equation}
where $d^a_c(k)$ gives the critical distance specific to the sub-set $\Omega^a_k$: for each space $\Omega^a$, the $k-$agents can afford the goods in the locations $X\in\Omega^a$
for which $D(X) \geq d^a_c(k)$. 
Similarly, a WTP
threshold $P_c^{*,a}$ can be defined for each \textit{arrondissement}. The agents with a WTP above the threshold of a given \textit{arrondissement}
can buy a good in any location in this \textit{arrondissement}. The WTP threshold, $P_c^*$, on the whole city is given by $P_c^*=max_a(P_c^{*,a})$.

\begin{figure}[!htbp]
\begin{tabular*}{\textwidth}{cc}
\includegraphics[width=0.47\textwidth]{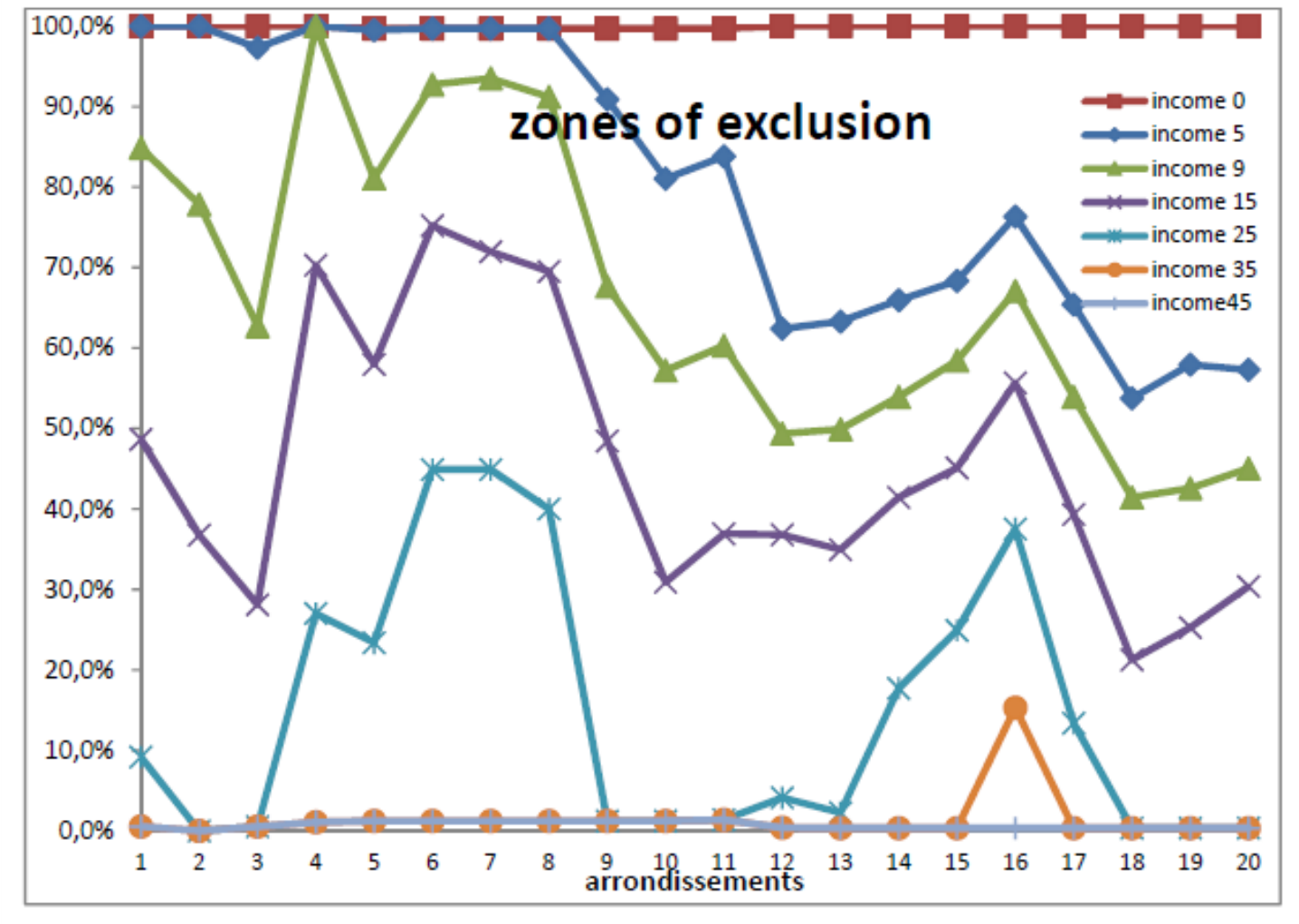} &
\includegraphics[width=0.47\textwidth]{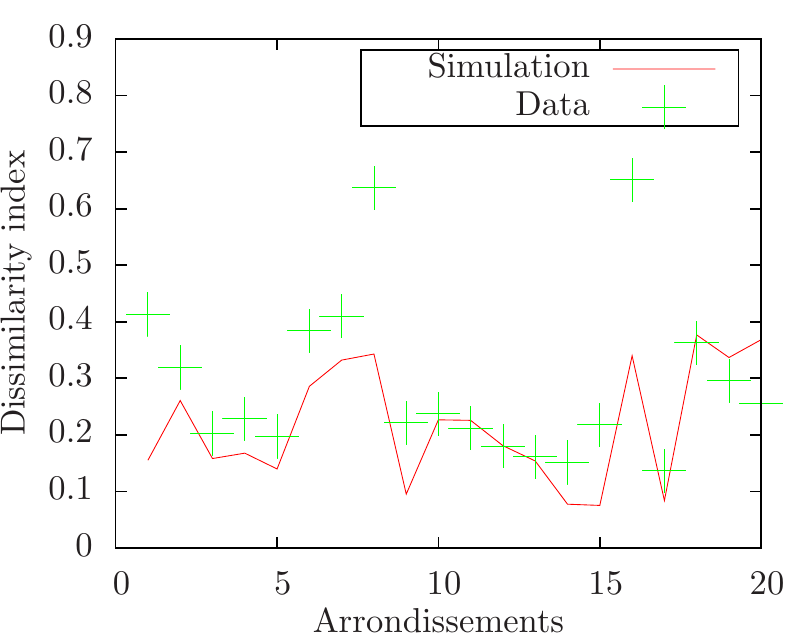}
\end{tabular*}
\caption{Left: for different WTP categories, the degree of exclusion from each arrondissement, measured by the percentage of arrondissement area where the considered category cannot buy a home (results from simulations with $K=50$). 
Right: social mix index computed for the empirical data (points) and simulations (curve).}
\label{exclusion}
\end{figure}

Figure \ref{exclusion} shows two different measures of the segregation within the (real and the simulated) city of Paris. On the left, we present the degrees of exclusion for different levels of income as obtained from simulations with $50$ income levels. Clearly, when the poorest people (income level $0$) are not able to buy a dwelling in Paris, the richest ones (income level $45$) can live just about wherever they want.
It is also interesting to note that, even for a low level of income (income level $9$ for example), there exist some opportunities to live in the expensive \textit{arrondissements}.
On the right-hand side of the figure, the social mix index (defined in Eq. \ref{eq:ID} in Section \ref{social_mix}) obtained in the simulations is compared with the one computed from real data, with transaction price being used as a proxy for income. The patterns are clearly similar, confirming the pertinence of our model for explaining prices in Paris.

In particular, the lowest levels of social diversity (corresponding to the highest values of the social mix index) are observed in the 8th,16th,18th,19th and 20th
\textit{arrondissements}, both in the simulations and in the data: the 8th and the 16th are the \textit{arrondissements} where people's incomes are above the threshold, while the 18th, 19th and 20th \textit{arrondissements} are the poorest.

\section{Conclusion}
\label{sec:conclusion}
Going beyond the simple Schelling segregation model, this paper studies social segregation through the spatial distribution of income resulting from dynamic price formation and agents' localization in a particular housing market model. In doing so, it actually introduces a
new general framework for studying such markets and the resulting
socio-spatial segregation. A specific feature of the model is the specification of the
attractiveness of each location, composed of both an
intrinsic part and a subjective part that depends on the agents' social preferences
and therefore evolves over time with
the social characteristics of each neighborhood.

For the particular choices made in the present study - a preference for neighbors with similar or higher incomes -,
the analysis of a non-saturated market yields the following main
results. First, as expected, for a wide range of parameters, a socio-spatial segregation
occurs, with richer people living in locations with higher intrinsic attractiveness.
Second, such segregation occurs only if social influence is strong enough: on its own, the heterogeneity in attractiveness is not enough to provoke segregation. Third, and more original, whenever socio-spatial segregation occurs, there exists a large area of locations with intermediate values of intrinsic attractiveness where some social diversity is preserved. Finally, we prove that segregation is due to the emergence of income thresholds which divide the space between richer and poorer people. This endogenous threshold varies following the dynamic attractiveness of the rich location. This result could have important economic implications for the regulation of the housing market. For example, a policy aiming to boost demand (by subsidizing the poorer potential buyers) without constraining the supply side would lead to an increase in the distribution of sellers' prices and thus exacerbate the segregation.

Specifying the results for the case of a monocentric city (the closer to the center, the higher the intrinsic attractiveness), all the results can be simply stated in terms of properties depending on the distance from the center. Whenever social influence is strong enough, there exists a critical distance isolating a central zone occupied by the richest people. Within this zone, there is no segregation between these rich agents: some are richer than others, and this shows up in the high variance in transaction prices. Poorer people find themselves on the outskirts, where social diversity is the weakest, leading to a low variance in transaction prices. At intermediate distances from the center, there is an area of moderate social mix.

Looking at the empirical distribution of prices in Paris for the year 1994, we observe that, as a first-order approximation, the simple
Alonso model (prices higher at the center) is a good description: higher prices are indeed concentrated around Notre Dame cathedral. The behavior of the mean and standard deviation of price transactions with the distance from the center are fairly well reproduced by the model. However, this monocentric characterization of the housing market in Paris is only correct on average; it misses certain particularities, such as the difference between prices on the left bank (higher) and the right bank (lower).
A better representation of the Paris housing market is
then proposed by considering an intrinsic attractiveness specific to
each of the 20 {\it arrondissements} of Paris.
Taking the spatial distribution of transaction prices as a proxy for the spatial distribution of income, the results compare favorably with empirical data in the distinction between \textit{arrondissements}: a low level of social diversity corresponds to \textit{arrondissements} with either the highest or the lowest average prices, and a lower level segregation corresponds to \textit{arrondissements} with intermediate average prices.

The mathematical analysis is limited to a non-saturated regime, as explained in Section \ref{sec:statio_gal}. On the technical side, this allows for a complete mathematical characterization of the stationary states. On the qualitative side, it demonstrates that heterogeneity in intrinsic attractiveness alone is not enough to provoke segregation, which only emerges when the social influence is strong enough.
Analytical study of the saturated regime - briefly explored numerically in Section \ref{sec:saturated} - will require a much more complex analysis, beyond the scope of the present paper. The present model provides a framework that is general enough to take into account other components of social preferences such as ethnic features or other population characteristics.

This study also allows for further investigations. Concerning social preferences, alternative hypotheses such as
agents preferring to live in more working-class areas have yet to be considered. Preliminary results indicate that a preference to live with poorer neighbors has opposite effects to those shown in the present paper, working against the emergence of segregation. Concerning the dynamics of the model, the long-term evolution will be studied by introducing a slow dynamics for the intrinsic attractiveness, corresponding to the effect of, e.g., investments in new amenities, or, on the contrary, the deterioration of a neighborhood. This should also allow for the analysis of the emergence of polycentric cities.
A further step would consist in analyzing the conditions on the
dynamics of subjective attractiveness for the emergence of ``hot
spots'' - local areas of high prices unrelated to the level of intrinsic attractiveness. This could accompany extensions to the model, notably in the line of \cite{Short} and \cite{Berestycki},
adding a diffusion term to the
attractiveness dynamics: high attractiveness of a location is
expected to have some positive influence on the attractiveness of
nearby locations. It is this kind of extension that is likely to lead to
dynamical instabilities, with the emergence of hot spots of high
prices,
that could, for example,
explain changes in the ranking of \textit{arrondissements} by property prices over the years. The study of these dynamical instabilities could provide new insights into the mechanisms behind gentrification.

\subsubsection*{Acknowledgements}
We are grateful to Marc Barth\'el\'emy and Jean Vannimenus for fruitful discussions at an early
stage of the present work -- and to Jean Vannimenus for the Russian proverb.
We thank the participants
of the Cambridge seminar on networks organized by Sanjeev Goyal for helpful remarks. Errors remain ours. We thank two anonymous referees for their helpful comments and advice, and Nicolas Bernigaud for the map shown in Fig. \ref{data-paris}.
This work is part of the project ``DyXi'' (in which all authors are involved)
supported by the  program SYSCOMM of the French National Research Agency, the ANR (grant ANR-08-SYSC-008). The authors thank the APUR ({\it Atelier Parisien d'URbanisme}, http://www.apur.org/)
for the data. Most of this work was done while LG was with the Laboratoire de Physique Statistique (LPS, Paris), being supported partly by a fellowship
allocated by the UPMC doctoral school ``ED389: Physics, from
Particles to Condensed Matter'', and partly by the DyXi grant.
LG is currently Junior Researcher at the ISI Foundation, Torino. AV is Associate Professor at the University Paris 2-Panth\'eon Assas. JPN is a CNRS member.

\bibliographystyle{model5-names}

\pagebreak

\renewcommand{\theequation}{A-\arabic{equation}} 
\setcounter{equation}{0}  
\renewcommand{\thesection}{A} 
\setcounter{section}{0}  

\section{Appendix}
\label{app:appendices}

\subsection{Proof of the Proposition, Section \ref{sec:prop}}
\label{app:Proof_prop1}
We give here the proof by recurrence on $k$. 

\textbf{Proof: case $k\geq\bar k$.}\\ 
At equilibrium,
$u^*_k(X)=\frac{v^*_k(X)}{\alpha}=\frac{\rho^*_k(X)}{\alpha}$ which gives:
\begin{equation}
 u^*_k(X)=\frac{\gamma}{K\alpha}\frac{A^*_k(X)}{(1/L^2)\,Z_k}
\label{equal}
\end{equation}
with
\begin{equation}
 Z_k\equiv\sum_{X\in\Omega} A_k^*(X)
\end{equation}
From the fixed point equations, we get, for any $k\in \{0,...,K-1\}$,
\begin{equation}
 Z_k= Z^0 \;+\; \frac{\epsilon}{\omega}\sum_{X\in\Omega} \sum_{k'\geq k} v_{k'}(X)=Z^0 \;+\; L^2\, \frac{\epsilon \gamma}{\omega K}\; (K-k)
\end{equation}
In particular, for the housed $(K-1)$-agents (highest WTP):
\begin{equation}
 u^*_{K-1}(X)= \frac{\gamma}{K\alpha}\frac{A^0(X)+\frac{\epsilon \alpha}{\omega }
 u^*_{K-1}(X)}{ (1/L^2) Z^0 +\frac{\epsilon}{\omega } \frac{\gamma}{K}}
\end{equation}
Solving this equation leads to:
\begin{equation}
 u^*_{K-1}(X)=\frac{\gamma}{K \alpha} \frac{A^0(X)}{(1/L^2) Z^0}
  \label{inst-stat}
\end{equation}
Now assume that (\ref{dens}) is true for $k+1\leq k'\leq K-1$ with $k\geq \bar{k}$. Then we have:
\begin{equation}
u^*_k(X)= \frac{\gamma}{K\alpha}\;\frac{A^0(X)+\frac{\epsilon \alpha}{\omega }u^*_k(X) + (K-k-1)\frac{\epsilon \gamma}{\omega K }\frac{A^0(X)}{(1/L^2)Z^0} }{(1/L^2) Z^0 +\frac{\epsilon \gamma}{\omega K }(K-k) }
\end{equation}
which gives $u^*_k(X)=\frac{\gamma}{K \alpha} \frac{A^0(X)}{(1/L^2)Z^0}$. Hence the proposition is proved by recurrence
for all $k\geq \bar{k}$.\\

\textbf{Proof: case $k < \bar k$ (assuming $\bar k >0$).}\\
From (\ref{rhot-stat}), the density of potential buyers in a location $X$
(outsiders) is given by:
\begin{equation}
\rho^*_k(X)=
\frac{A^*_k(X)}{\sum_{X'\in \Omega}A^*_k(X')}\;\sum_{X'\in \Omega} \rho^*_k(X')
 \label{rhok}
\end{equation}

For $k<\bar{k}$, $v^*_k(X)=0$ for $X \in \Omega\setminus\Omega_k$, and $\bar{v}^*_k(X)=0$ for $X\in \Omega_k$.
Hence, since $\rho^*_k(X)=v^*_k(X)+\bar{v}^*_k(X)$, we have:
\begin{equation}
\sum_{X\in \Omega_k}\rho^*_k(X)=\sum_{X\in \Omega_k}v^*_k(X)=\sum_{X\in \Omega} v^*_k(X)
\end{equation}
As we have seen, in the stationary state the total number of  $k$-agents
on the lattice is $\frac{\gamma L^2}{\alpha K}$, and $v^*_k(X)=\alpha u^*_k(X)$, so that finally:
\begin{equation}
\sum_{X\in \Omega_k}\rho^*_k(X)=L^2\,\frac{\gamma}{K}
 \label{rhotot}
\end{equation}
Then, summing Eq. (\ref{rhok}) over $X$ in $\Omega_k$, we get:
\begin{equation}
 \rho^*_k(X)=\frac{\gamma L^2}{K}\frac{A^*_k(X)}{\sum_{X'\in \Omega_k} A^*_k(X')}
\end{equation}
Let us now consider $k=\bar{k} -1$, that is the highest WTP category for which $\Omega_k$ is different from $\Omega$, i.e., the highest $k-$WTP which does not allow to buy a good anywhere.
The agents with a WTP greater than $P_k$ are distributed according to the expression (\ref{dens}). Thus, if $X\in \Omega_k$:
\begin{eqnarray}
\rho^*_k(X) & = &v^*_k(X) = \alpha u^*_k(X)=\frac{\gamma L^2}{K}\frac{A^*_k(X)}{\sum_{X'\in \Omega_k} A^*_k(X')}\\
 u^*_k(X) & =  &\frac{\gamma L^2}{K\alpha}\frac{A^0(X)+\frac{\epsilon \alpha}{\omega }[(K-k)\frac{L^2\,A^0(X)}{\sum_{X'\in \Omega A^0(X')}} +u^*_{k}(X)]}
 { \sum_{X'\in \Omega_k}[A^0(X')]+\frac{\epsilon}{\omega} [ (K-k)\frac{\gamma L^2}{K}\frac{\sum_{X'\in \Omega_k} A^0(X')}{\sum_{X'\in \Omega} A^0(X')}+\frac{\gamma L^2}{K}]}
\label{eq-inst}
\end{eqnarray}
The equation (\ref{eq-inst}) leads to:
\begin{equation}
 u^*_k(X)=\frac{\gamma L^2}{K \alpha} \frac{A^0(X)}{\sum_{X'\in \Omega_k} A^0(X')}.
\end{equation}
By recurrence, we can generalize to any $k<\bar{k}$:
\begin{eqnarray}
 v^*_k(X)  =  \alpha u^*_k(X) & = &\frac{\gamma L^2}{K} \frac{A^0(X)}{\sum_{X'\in \Omega_k} A^0(X')} \textrm{ if $X\in \Omega_k$ } \\
 & = & 0 \textrm{ otherwise}
\end{eqnarray}

Note that the above proof can easily be shown to apply to the case where agents weight higher categories differently, as in Eq.~(\ref{wvk>}).

\subsection{Computing the mean attractiveness}
\label{sec:Abar}
Assuming $\bar{k} \in\{1,...,K-1\}$, we obtain the expression of $\bar{A}^*(X)$ as follows.
From the stationary equation for $A_k^*$, Eq.(\ref{A-stat}), and from the Proposition stated in Section \ref{sec:prop}, we get, in a non-saturated equilibrium:
\begin{equation}
A_k^*(X)=A^0(X)+ \frac{\epsilon \gamma}{\omega K} \sum_{k'\geq \max(k,k_c(X))}\;\frac{A^0(X)}{\langle A^0  \rangle_{k'}}
\label{ak*}
\end{equation}
$k_c(X)$ is the smallest value of $k$ such that the $k$-agents are present on $X$.

First consider $k\geq \bar{k}$. In this case, $\langle A^0  \rangle_k = \langle A^0  \rangle$ so that, from (\ref{ak*}), we get:
$A^*_k(X)=A^0(X)+ \frac{\eta\; A^0(X)}{2 }[1 - \frac{k}{K}]$,
where $\eta=\frac{\epsilon \gamma}{\omega \, \langle A^0 \rangle}$, as defined in (\ref{eta}).
At locations $X$ where $k_c(X)=\bar{k}$, for $k < \bar{k}$, we obtain from (\ref{ak*})
$A^*_k(X)=A^0(X)+ \frac{\eta\; A^0(X)}{2}[1 -\frac{\bar{k}}{K}].$
As a result, the average value $\bar{A}^*(X)=\sum_0^{K-1} A^*_k(X) $ is
\begin{equation}
\bar{A}^*(X)=A^0(X)+ \frac{\eta\; A^0(X)}{2}[\frac{K+1}{K}-\frac{(\bar{k}+1)\bar{k}}{K^2}]
\end{equation}

\subsection{The residential city}
\label{sec:living}
The densities obtained with the partial differential equations approach can take arbitrarily small values. In the agent-based version, the number of agents at a given location obviously only take integer values: we can expect a good comparison considering that
the actual density is zero whenever $a^2\, u_k(X)$ is less than $1$. For a concrete example, consider a monocentric city with $A^0(X)=A^0_{max} \exp(-D(X)^2/R^2)$, with $R<{\cal D}$.
The diameter ${\cal D}_{eff}$ of the space of residential locations is obtained by writing $a^2\, \sum_k u_k(X)=1$. Taking as an approximation the expression (\ref{dens}) for every location, and setting a continuous limit by assuming $a$ small, we obtain how ${\cal D}_{eff}$ (or equivalently, the number of active sites,  $L_{eff} ={\cal D}_{eff} / a$) scales with the model parameters:
\begin{equation}
{\cal D}_{eff} = R \sqrt{ \ln ( \frac{{\cal D}}{R}\, \sqrt{\frac{\Gamma}{2\pi\alpha}})}
\label{Deff}
\end{equation}
Note that the demand pressure, expressed by the ratio $\frac{\Gamma}{\alpha}$, only contributes through a logarithmic term. For the parameter values (notably ${\cal D} = a L=100$, $R=10$) used in the numerical simulations in Section \ref{sec:num}, the above formula gives ${\cal D}_{eff} \simeq 28$, in keeping with what can be seen in, e.g., Fig.~\ref{occup}.

\subsection{Extension to an arbitrary WTP distribution} 
\label{app:gen}

In the model we have assumed that the $K$ values of WTP are uniformly distributed among the agents in the external reservoir. One may want to start with
a particular probability distribution function (PDF) of the reservation price, corresponding to the empirical income distribution of a given population, for example.
Let $Q(P)$ be this PDF, with support on $[P_0, P_0+\Delta ]$. We can construct the discrete set of $K$ levels from  $Q(P)$ by requiring the same density of agents in any one of these levels. The $\{P_k, k=0,...,K-1\}$ are thus defined by:
\begin{equation}
\int_{P_0}^{P_{k}}\; Q(p)dp \;=\; \frac{k}{K}
 \label{histoequal}
\end{equation}
for $k=1,...,K$ (with $P_K \equiv P_0+\Delta$).
Of particular interest is the large $K$ limit. In this case, we have $P_k=P(k/K)$, the function $P(z)$ on $z\in [0, 1]$ being given as solution of $\int_{P_0}^{P(z)}\; Q(p)dp \;=\; z$, or, equivalently by:
\begin{equation}
Q(P(z))\;P'(z)=1
 \label{histoequal_largeK}
\end{equation}
where $P'(z)=dP/dz$, with $P(0)=P_0$ and $P(1)=P_0+\Delta$.\\

Formally, all the analysis done in this paper for the uniform case, (\ref{even}), can be applied to an arbitrary WTP function.
In particular, the critical WTP threshold is here
 given by $P_c^*=P(z_c^*)$, $z_c^*$ being obtained by solving the equation
\begin{equation}
\frac{1}{\widetilde \xi}\; \log \frac{P(z_c^*)}{P^0} \,-\,1 \;=\;  \frac{\eta}{2} \left(1 - \left(z_c^*\right)^2 \right).
\label{eq:Pcbis}
\end{equation}

The above results can be generalized to the case of agents weighting higher categories differently, as in Eq.~(\ref{wvk>}), with $w_{k'-k}=w(\frac{k'-k}{K})$. We get:
\begin{equation}
\frac{1}{\widetilde \xi}\; \log \frac{P(z_c^*)}{P^0} \,-\,1 \;=\;\eta\;  \int_{z_c^*}^1 dt'\,\int_0^{t'}dt\; w(t'-t).
 \label{eq:Pcter}
\end{equation}
The critical value $\eta_c$ of $\eta$ is obtained by setting $z_c^*=0$ in the above equation.

\end{document}